\RequirePackage{fix-cm}
\documentclass[twocolumn,epjc3]{svjour3}  
\smartqed  
\RequirePackage{graphicx}
\journalname{Noname}

\usepackage{multirow}
\usepackage{subfigure}
\usepackage{bm}
\usepackage{bbm}
\usepackage{braket}
\usepackage{lineno}
\usepackage{slashed}
\usepackage{amsmath}
\usepackage{amssymb}
\usepackage{amsfonts}
\usepackage{hyperref}
\usepackage{microtype}
\hypersetup{colorlinks=true,linkcolor=red,filecolor=black, citecolor=blue}

\usepackage[
    natbib=true,
    style=phys,
    biblabel=brackets,
    giveninits=true,
    abbreviate=false,
    doi=false,
    url=false,
    isbn=false,
    block=space,
    backref=false,
    backrefstyle=two,
]{biblatex}
\usepackage{csquotes}
\begin{document}

\title{Podolsky quantum electrodynamics for strongly coupled Dirac fermions in (2+1)D
}


\author{Carlos A. P. C. Junior\thanksref{e1,addr1}
        \and
        Leandro O. Nascimento\thanksref{e2,addr2,addr3} \and
        Van Sérgio Alves\thanksref{e3,addr1,addr2}
}

\thankstext{e1}{e-mail: carlos.chaves.junior@icen.ufpa.br}
\thankstext{e2}{e-mail: lon@ufpa.br}
\thankstext{e3}{e-mail: vansergi@ufpa.br}


\institute{Programa de P\'os-Graduação em Física, Universidade Federal do Par\'a, 66075-110, Bel\'em, Par\'a, Brazil \label{addr1}
           \and
           Faculdade de F\'isica, Universidade Federal do Par\'a, 66075-110, Bel\'em, Par\'a, Brazil \label{addr2}
           \and
           Programa de P\'os Graduação em Física, Universidade Federal de Campina Grande, 58429-900, Campina Grande, Para\'iba, Brazil\label{addr3}
}

\date{Received: date / Accepted: date}

\maketitle

\begin{abstract}
We investigate generalized quantum electrodynamics (GQED), a higher-derivative extension of quantum electrodynamics in (3+1) dimensions. We perform a dimensional reduction of this theory to (2+1)D by confining the Dirac current to a plane while allowing the gauge-field to propagate outside the plane. The resulting effective theory, which we denote as Pseudo Generalized QED (PGQED), is minimally coupled to massless Dirac fermions. At strong coupling, we show that a dynamical mass is generated from approximate solutions to the Schwinger-Dyson equations, leading to dynamical chiral symmetry breaking and modifications of the fermion dispersion relation. We derive an analytic critical coupling constant $\alpha_{c}(\mu)$ and a critical flavor number $N_{c}(\mu)$, which depend on the Podolsky parameter $\mu$ and the ultraviolet cutoff $\Lambda$. These analytical results are in good agreement with our numerical solutions. Finally, we discuss how the model may provide qualitative insights for graphene, estimating a viable range for $\mu$ in the ultrarelativistic limit and highlighting possible applications to two-dimensional materials.
\end{abstract}

\section{Introduction}
\label{intro}

One of the most important features of low-dimensional quantum field theories is the possibility of chiral symmetry breaking in (2+1)D for both QED and four-fermion interacting theories (such as the Nambu-Jona-Lasinio model) \cite{Appelquist1985,Alves2013}. The order parameter for chiral symmetry breaking is the so-called dynamically generated mass of the fermion, usually calculated using the Schwinger-Dyson equations. Interest in this subject first emerged as a simplified framework for studying confinement in quantum chromodynamics in the strongly correlated limit \cite{Roberts1994}. Following the experimental realization of graphene \cite{Novoselov2005} and the observation of Dirac cones in graphene, these effective models have also been applied to the physics of two-dimensional materials \cite{Alves2013,CastroNeto2009,Gusynin2007}. In this case, the main idea is to derive a quantum phase transition from a previously massless spectrum to an electronic band with a gap, which has direct consequences for transport properties of the material \cite{CastroNeto2009}.

When considering the physics of two-dimensional materials, Pseudo quantum electrodynamics (PQED) \cite{Marino1993} has successfully described several phenomena that have been compared with experimental findings \cite{Silva2017,Silva2024,Fernndez2020,Fernndez2021,Bezerra2023,Menezes2017}. PQED is obtained through a dimensional reduction of QED in (3+1)D, where the main physical assumption is that the matter current is confined to a spatial plane. Furthermore, the dimensional reduction of QED with a massive photon has also been explored in Ref.~\cite{Magalhes2021}, where an effective gauge-field theory describing the Yukawa interaction in (2+1)D is derived. Emergent gauge-fields, have been shown to be an important tool for calculating interacting effects among either confined electrons or quasiparticles in two-dimensional materials. Within many-body physics, the role of the Yukawa potential is to modify the Coulomb potential such that the screening of charges becomes relevant for the physics of the system. Similar ideas have also been explored using ultracold atoms in optical lattices \cite{Magalhes2021}.

The GQED, proposed by Boris Podolsky in 1942, is a model that introduces a higher-derivative term into standard quantum electrodynamics (QED), yielding a characteristic parameter $a$ with units of inverse mass, called the Podolsky parameter \cite{Podolsky1944}. Whenever this parameter vanishes, one recovers Maxwell's theory. This GQED preserves both Lorentz and gauge symmetries while describing an additional massive gauge mode, beyond the standard massless photon \cite{Fonseca2010}. Furthermore, GQED provides a finite potential for a point-like particle at short distances, unlike electrons within conventional QED \cite{Podolsky1944}. Hence, the model exhibits an improved ultraviolet behavior of the model in comparison with QED, which may have consequences for the renormalizability of the theory. Recently, GQED has been investigated from several perspectives, including gauge fixing and Pauli–Villars regularization~\cite{Ji2019}, canonical and path-integral quantization~\cite{Thibes2017}, and light-front dynamics~\cite{Bertin2025}. Nevertheless, the dimensional reduction of GQED has not been considered until now. In this work we address this gap by deriving PGQED and demonstrating its improved ultraviolet behavior compared with PQED. In Ref.~\cite{Neves2025}, a similar model has been considered and its Higgs mechanism and unitarity at tree level have been discussed.

In this work, we perform a dimensional reduction of GQED and obtain a reduced model that we call PGQED, fully defined in (2+1)D. Within the strongly correlated limit, we calculate a dynamically generated mass for an initially massless Dirac electron. As in PQED, we consider that the matter current is confined to move only in a spatial plane ($j_z=0$), such as electrons in a monolayer of graphene, while the gauge-field is free to propagate in (3+1)D. This hybrid situation has been discussed in Ref.~\cite{Marino1993} in the context of QED, which serves here primarily as a physical motivation for the class of reduced gauge theories considered, rather than as a direct target for quantitative comparison. For PGQED, the static potential among electrons reads $V(r)\propto (1-e^{-r/a})/r$, recovering the Coulomb interaction when $a\rightarrow 0$. This kind of interaction allows us to approximate the static electromagnetic interaction by a local four-fermion action, given by $\propto \rho_0^2/a$ at short distances $r\ll a$, where $\rho_0=\bar\psi\gamma_0\psi$ is the charge density of the Dirac field $\psi$. For a full description of the interaction, as will become clear later, we formulate the theory in terms of a pseudo-differential gauge operator. 

Throughout this work, PGQED is treated as an effective quantum field-theory framework designed to isolate the impact of higher-derivative gauge dynamics on dynamical mass generation. Our analysis is not intended to provide a complete microscopic description of graphene or other specific materials, but rather to identify universal strong-coupling features that may later be incorporated into more realistic models.

The dynamical mass generation implies chiral symmetry breaking in (2+1)D whenever we consider Dirac fermions in the four-dimensional representation of the Dirac matrices \cite{Appelquist1985,Roberts1994}. Hence, for strong interactions, we employ a suitable set of approximations to solve the Schwinger-Dyson equations. In particular, we obtain nontrivial solutions for the mass function $\Sigma(p)$ of the full electron propagator using both the rainbow-quenched and rainbow-unquenched approximations. In general, when $\Sigma(p)\neq 0$, we have a dynamically generated mass. In the particular case $\Sigma(p\rightarrow 0)=M$, it follows that $M$ is the pole of the full fermion propagator and, therefore, can be interpreted as the dynamical mass of the electron. Furthermore, we show that $M$ corresponds to a small fraction of the UV cutoff $\Lambda$, which sets the effective validity scale of the theory. This parameter can be related to the lattice constant $a_L\approx 10^{-10}\,\mathrm{m}$ of graphene; hence, $\Lambda\approx 1\,\mathrm{eV}$ serves as an upper energy limit for the Dirac cones \cite{Geim2011}. Indeed, for energies above this value, one expects corrections to the Dirac cones and, therefore, the Dirac approximation for electrons in graphene becomes less reliable \cite{CastroNeto2009}. Throughout this work, $\Lambda$ is regarded as the ultraviolet validity scale of effective theory. We also consider a full Lorentz invariant matter field, which simplifies the calculation of the electron self-energy. This can also be realized in graphene in the low-density limit of charge carriers, as discussed in Ref.~\cite{Geim2011}. In this context, our model parameters $(\Lambda, \mu, \alpha)$ can be estimated from experimental data only at the level of order-of-magnitude guidance, rather than as quantitative predictions, since PGQED is not intended as a microscopic description of graphene. Improved approximations show that there exists a critical coupling constant $\alpha_c$ and a critical flavor number $N_c$ that separate the symmetric and broken phases. These parameters constrain the values of the fine-structure constant $\alpha\propto e^2/4\pi$ ($\alpha>\alpha_c$) and the number of fermion flavors $N$ ($N<N_c$). Finally, we describe the role of the Podolsky parameter $\mu=1/a$ in driving this quantum phase transition.

To improve the reliability of our results beyond the standard rainbow-ladder approximation, we also incorporate the Ball-Chiu vertex into the Schwinger-Dyson formalism. Unlike the bare interaction vertex, the Ball-Chiu construction identically satisfies the Ward-Takahashi identity, thereby preserving the fundamental gauge invariance of the theory at the non-perturbative level \cite{Fischer2004}. This approach introduces a non-trivial momentum dependence to the fermion wave-function renormalization $A(p)$, leading to a complex dynamical interplay between the vertex enhancement in the infrared regime and the screening effects governed by the Podolsky parameter $a=1/\mu$. Evaluating this coupled system allows us to assess how higher-order, gauge-invariant corrections quantitatively impact the critical coupling constant and the dynamically generated mass.

The outline of this paper is as follows: In Sec.~II, we derive our model and its Feynman rules. In Sec.~III, we discuss the Schwinger-Dyson equations for our model. In Sec.~IV and V, we calculate the mass function using a suitable set of approximations. In Sec.~VI, we calculate numerically the mass function and the fermion wave-function renormalization using the Ball-Chiu vertex. In Sec.~VII, we discuss our main results and their consequences. We have included three appendices:  \ref{A}, where we show details about the dynamical mass generation; \ref{B}, where we present the numerical results for the mass function and the wave-function renormalization of the electron;  \ref{C}, where we consider the fact that electrons propagate with the Fermi velocity in graphene.

\section{The Model}
\label{model}

In this section, we derive the dimensional reduction of GQED from (3+1)D to (2+1)D \cite{Ortega2018}. The derivation follows the same general procedure employed in the construction of PQED \cite{Marino1993}.

We begin with the GQED Lagrangian in Euclidean space,
\begin{eqnarray}\label{QEDPod}
    \mathcal{L}_{{\rm GQED}}=\frac{1}{4}F_{\mu\nu}F^{\mu\nu}-\frac{a^2}{2}\partial_\nu F^{\mu\nu}\partial^\sigma F_{\mu\sigma}+\nonumber\\
    +\bar{\psi}(i\slashed{\partial}-m)\psi+e\bar\psi\gamma_\mu \psi A^\mu+\frac{\xi}{2}(\partial_\mu A^\mu)^2,
\end{eqnarray}
where $A_\mu$ is a gauge-field and $F_{\mu\nu}=\partial_\mu A_\nu-\partial_\nu A_\mu$ is its field-strength tensor. The constant $a$ is the Podolsky parameter, which multiplies a higher-derivative term. We also include the gauge-fixing parameter $\xi$. On the other hand, $\psi$ is the Dirac field, $m$ is its bare mass, $e$ is the elementary electric charge, and $\gamma_\mu$ are the Dirac matrices with $\mu=0,1,2,3$. We also consider $c=\hbar=1$ everywhere. The details regarding the dimensional reduction are shown in  \ref{A}. From these calculations, it follows that the reduced version of Eq. \eqref{QEDPod} is
\begin{eqnarray} \label{PGQED}
    \mathcal{L}_{{\rm PGQED}}=\frac{1}{4}\Tilde{F}_{\mu\nu}K_E[\square]\Tilde{F}^{\mu\nu}+\bar{\psi}(i\slashed{\partial}-m)\psi+\nonumber\\
    +e\bar\psi\gamma_\mu \psi \Tilde{A}^\mu+\frac{\xi}{2}\partial_\mu\Tilde{A}^{\mu}K_E[\square]\partial_\nu \Tilde{A}^\nu,
\end{eqnarray}
where $\Tilde{A}_{\mu}$ is fully defined in (2+1)D ($\mu=0,1,2$) and $\Tilde{F}_{\mu\nu}=\partial_\mu \Tilde A_\nu-\partial_\nu \Tilde A_\mu$ is its field-strength tensor, similarly to Eq.~(\ref{QEDPod}). We refer to the reduced theory defined in Eq.(\ref{PGQED}) as Pseudo Generalized Quantum Electrodynamics (PGQED). A similar calculation has been discussed in Ref.~\cite{Neves2025}, using a different kind of higher-order derivative term. 
 
\begin{figure}
    \centering
    \includegraphics[width=\linewidth]{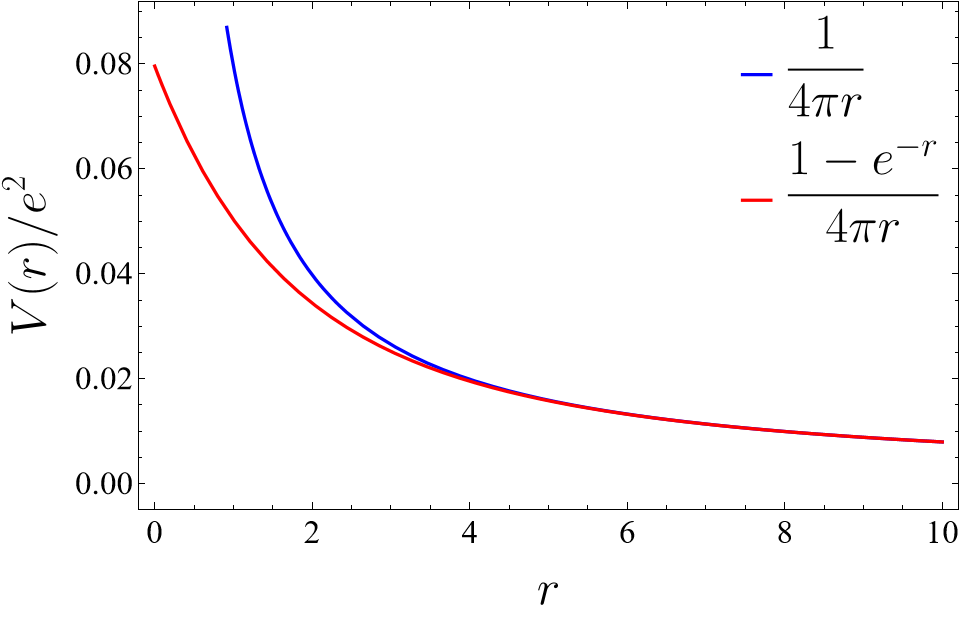}
    \caption{\textit{The static potential of PGQED}. The red curve shows the PGQED static potential given by Eq.~\eqref{Staticpot} for $a=1$. The blue curve corresponds to the Coulomb potential recovered in the limit $a\rightarrow \infty$. Unlike the Coulomb potential, the PGQED potential remains finite at the origin.}
    \label{coulombvspod}
\end{figure}

Furthermore, the pseudo-differential operator $K_E[\square]$ reads
\begin{equation} \label{KernelRGQED}
    K_E[\Box]=\frac{2(-\square+\mu^2)}{(-\square+\mu^2)(-\square)^{1/2}-(-\square)(-\square+\mu^2)^{1/2}},
\end{equation}  
where we have defined a new constant $\mu=1/a$ with dimensions of mass and $\square$ is the standard d'Alembertian operator. This nonlocal kernel reproduces the interaction mediated by the original four-dimensional theory while describing electrons confined to a two-dimensional plane. We now consider the interaction between static charges. The dimensional reduction is constructed so as to preserve the electron-electron interaction (see \ref{A}). Consequently, one recovers the same static potential of GQED in Eq.~(\ref{QEDPod}), namely,
\begin{equation} \label{Staticpot}
     V(r)=\frac{e^2}{4\pi r}(1-e^{-r/a}).
\end{equation}
For small distances $r\ll a$, the potential is finite and is given by $e^2/4\pi a$ at lowest order, which is a feature of GQED. For larger distances $r\gg a$, we obtain the Coulomb potential as in QED in (3+1)D. On the other hand, for intermediate values $r\approx a$, we have a Coulomb potential with a screened electric charge. In Fig.~\ref{coulombvspod}, we plot Eq.~(\ref{Staticpot}) for an illustrative value of  $a$; all units are arbitrary. These results mainly refer to the classical aspects of GQED. In the following, we shall consider the quantum effects of the dimensional reduction of GQED.

In the realm of quantum field theory, in particular for QED in (3+1)D, we need to deal with divergences when calculating quantum corrections to the Green functions of the model. It turns out that the improved behavior as $r\rightarrow 0$ in GQED generally provides a mechanism for dealing with these divergences. For our model, given by Eq.~(\ref{PGQED}), we should also observe a better UV behavior compared to PQED, which can be analyzed by calculating the Feynman rules of this action.

\subsection{The Feynman Rules}

In this subsection, we derive the tree-level Feynman rules for PGQED in Eq.~(\ref{PGQED}). These rules are required for the formulation of the Schwinger-Dyson equations used to determine the dressed Green's functions, which we shall discuss later. After straightforward algebra, we obtain the bare gauge-field propagator, given by
\begin{equation}\label{pppqed}
    \Delta^0_{\mu\nu}(p)=\left[\frac{1}{2}\frac{1}{(p^2)^{1/2}}-\frac{1}{2}\frac{1}{(p^2+\mu^2)^{1/2}}\right]\left[\delta_{\mu\nu}-\frac{p_\mu p_\nu}{p^2}\right],
\end{equation}
where throughout this work we adopt the Landau gauge ($\xi\to+\infty$). Integrating out the gauge-field in Eq.~(\ref{PGQED}), it is straightforward to conclude that static charges interact through the potential $V(r)$ in Eq.~(\ref{Staticpot}), which is obtained from the Fourier transform of $\Delta^0_{00}(p_0=0,\textbf{p})$ \cite{Marino2017}. In the limit $\mu\rightarrow\infty$ (equivalently, $a\rightarrow 0$), we recover PQED, as expected. Unlike standard GQED, the square-root structure of the propagator (see Eq.~(\ref{pppqed})) prevents its interpretation in terms of isolated massive or massless particle poles, since its analytic structure is characterized instead by a branch cut. Its classical analytic properties have been discussed in Ref.~\cite{Neves2025}.

On the other hand, the bare fermion propagator reads
\begin{equation} \label{Fferprop}
    S_{0F}=\frac{1}{-\slashed{p}+m}
\end{equation}
and the bare interaction vertex is given by
\begin{equation}
    e\Gamma^\mu_0=-e\gamma^\mu,
\end{equation}
where the Dirac matrices $\gamma^\mu$ satisfy the Clifford algebra, namely, $\{\gamma^\mu,\gamma^\nu\}=-2\delta^{\mu\nu}$. From now on, let us assume that these matrices are written in a four-dimensional representation, which allows us to discuss chiral symmetry breaking driven by a dynamically generated mass \cite{Appelquist1985,Alves2013}. From Eq.~(\ref{Fferprop}), we obtain, after a Wick rotation $p_0\equiv E\rightarrow i E$, the usual Dirac dispersion relation $E_\pm(\textbf{p})=\pm\sqrt{\textbf{p}^2+m^2}$. The model is invariant under chiral transformations in the limit $m\rightarrow 0$. Interestingly, in both QED and PQED this symmetry may be broken by quantum corrections in the strong-coupling regime. Here, we investigate how the Podolsky parameter $\mu$ modifies the dynamics of chiral symmetry breaking. In order to do so, we shall define the Schwinger-Dyson equations for the full propagators.

\section{Schwinger-Dyson Equations for PGQED}
\label{SchwingerDysonPGQED}

A dynamically generated mass for the Dirac field implies chiral symmetry breaking. This phenomenon is intrinsically non-perturbative in quantum field theory. Here, we use the Schwinger-Dyson equations for PGQED in order to study this problem.

These equations form a set of coupled nonlinear integral equations involving all the dressed Green's functions of a given theory and, once solved, provide access to the full physics described by the model \cite{Roberts1994,Maris1995,Maris1996}.

\begin{figure}
    \centering
    \includegraphics[width=\linewidth]{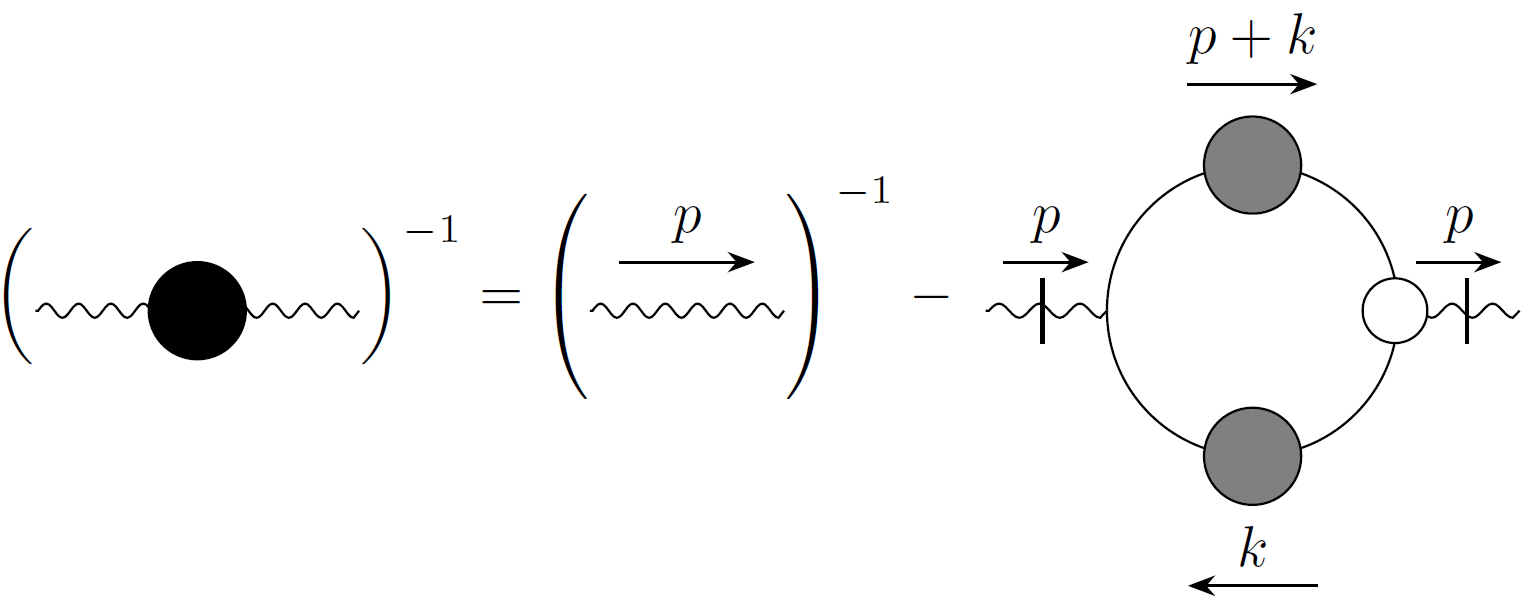}
    \caption{\textit{The Schwinger-Dyson equation for the gauge-field propagator in Eq.~(\ref{esdpp})}. Filled dots represent full propagators and vertex. The black dot corresponds to the full gauge-field propagator, the gray dot to the full fermion propagator, and the white dot to the vertex. The first term on the right-hand side is the bare gauge-field propagator, while the second term represents the gauge-field self-energy $\Pi^{\mu\nu}(p)$.}
    \label{esdgf}
\end{figure}

Since we are interested in calculating the dynamically generated mass of the fermion, it suffices to determine the full electron propagator. Therefore, let us consider only the full two-point functions for both the gauge and matter fields, which are coupled to the full three-point vertex interaction. Hence, we need the Schwinger-Dyson equations for the three relevant Green functions of this problem: one equation for the gauge-field propagator; an equation for the fermion propagator; and one equation for the vertex interaction \cite{Roberts1994}. In the case of the gauge-field propagator, we have
\begin{equation}\label{esdpp}
    \Delta^{-1}_{\mu\nu}(p)=\Delta^{-1}_{0\mu\nu}(p)-\Pi_{\mu\nu}(p),
\end{equation}
where $\Delta^{-1}_{\mu\nu}(p)$ is the full gauge-field propagator with all of the radiative corrections. The first term on the rhs of Eq.~(\ref{esdpp}) is the inverse of the gauge-field propagator at tree level, given by Eq.~(\ref{pppqed}), and the second term is the gauge-field self-energy, given by
\begin{equation} \label{PiGF}
    \Pi^{\mu\nu}(p)=-e^2\int\frac{d^{3}k}{(2\pi)^3}\text{Tr}\{\Gamma^\mu_0 S_F(k+ p)\Gamma^\nu(k,p)S_F(k)\},
\end{equation}
where $\Gamma^\nu(k,p)$ is the full vertex interaction, and $S_F(p)$ is the  full electron propagator. Interestingly, within perturbation theory, the gauge-field self-energy $\Pi_{\mu\nu}(p)$ is given by the vacuum-polarization tensor at lowest order. In Fig.~\ref{esdgf}, we show the diagrammatic representation of Eq.~(\ref{PiGF}). 

\begin{figure}
    \centering
    \includegraphics[width=\linewidth]{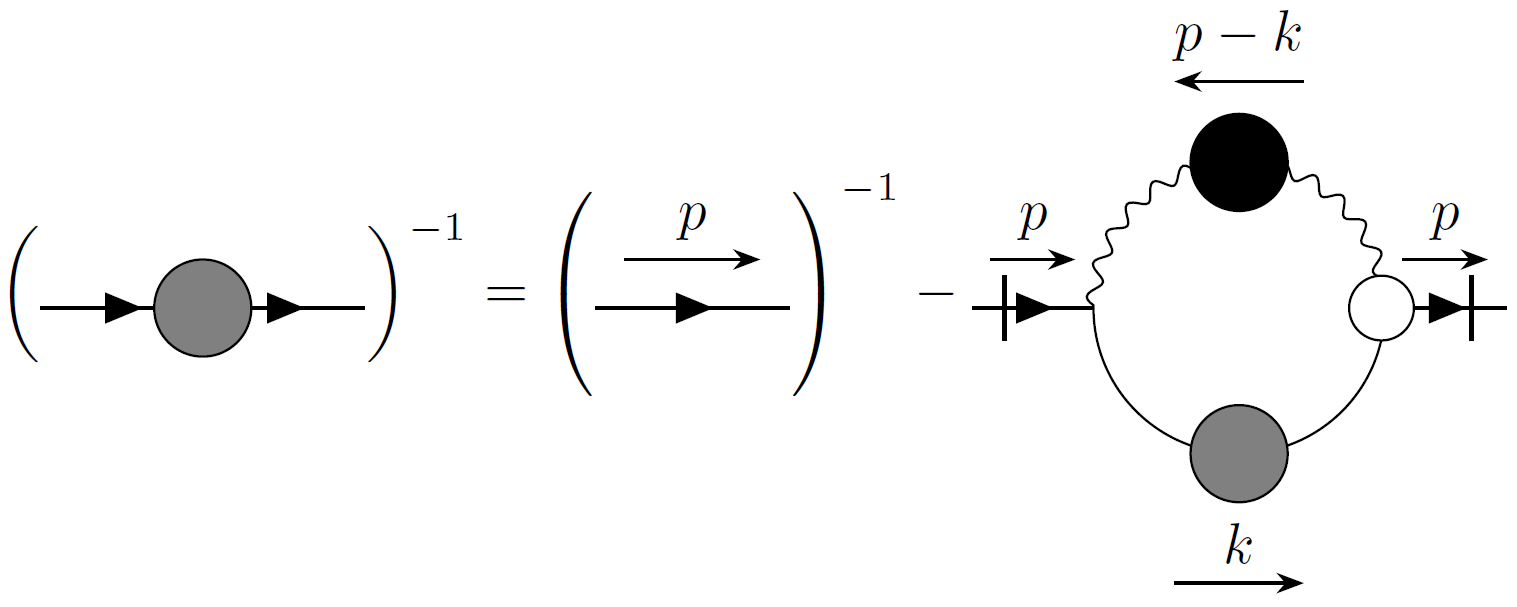}
    \caption{\textit{The Schwinger-Dyson equation for the full electron propagator in Eq.~(\ref{SDfer})}. Filled dots represent full propagators and vertex. The black dot corresponds to the full gauge-field propagator, the gray dot to the full fermion propagator, and the white dot to the vertex. The first term on the right-hand side is the bare fermion propagator and the second term represents the fermion self-energy $\Xi(p)$.}
    \label{esdf}
\end{figure}

For the fermion propagator, the Schwinger-Dyson equation reads
\begin{equation}\label{SDfer}
    S^{-1}_F(p)=S^{-1}_{0F}(p)-\Xi(p),
\end{equation}
where
\begin{equation}\label{sdeff}
    \Xi(p)=e^2\int\frac{d^{3}k}{(2\pi)^3}\Gamma^\mu_0 S_F(k)\Gamma^{\nu}(k,p )\Delta_{\mu\nu}(p-k)
\end{equation}
is the electron self-energy. From Eq.~\eqref{sdeff}, we conclude that, in order to calculate $\Xi(p)$, we need to know the full gauge-field propagator, which also depends on the electron propagator through Eq.~(\ref{PiGF}). We show the diagrammatic representation of this equation in Fig. \ref{esdf}. The coupled structure of the Schwinger-Dyson equations requires a suitable truncation scheme in order to determine the full fermion propagator.

The remaining Schwinger-Dyson equation governs the full vertex function~\cite{Roberts1994}, however, in this work we will consider the rainbow approximation. Within this approximation, the full vertex is replaced by its bare counterpart, which greatly simplifies the vertex function by retaining only the bare term given by the Dirac matrices \cite{Maris1996}. Next, within the called quenched approximation, the fermionic loops are neglected, hence, $\Pi^{\mu\nu}\rightarrow 0$, which yields $\Delta_{\mu\nu}\rightarrow\Delta^0_{\mu\nu}$ by using Eq.~(\ref{esdpp}). This set of approximations is known as the rainbow-quenched approximation and it is a valuable way for calculating dynamical mass generation. A systematic improvement is obtained by incorporating a gauge-field self-energy calculated within the large-$N$ approximation, known as the rainbow-unquenched approximation. In principle, this provides better results as long as the condition $N\gg 1$ is satisfied by the physical system. Several works have confirmed the validity of these approximations after including several corrections, such as a better vertex function, lattice theory, and improved numerical tools \cite{Maris1996,Roberts1994}. Before introducing these approximations, we parameterize the full fermion propagator in terms of two scalar functions, i.e.,
\begin{equation}\label{swdfe}
    S_F^{-1}(p)=-A(p)\slashed{p}+\Sigma(p),
\end{equation}
where $A(p)$ is the fermion wave-function renormalization and $\Sigma(p)$ is called the mass function. The main motivation for the \textit{ansatz} in Eq.~(\ref{swdfe}) is that it reproduces all Lorentz-invariant terms that originally appear in the bare electron propagator. Furthermore, it allows $\Sigma(p)\neq 0$ even when $m\rightarrow 0$, which cannot be realized by a perturbative expansion in PGQED. $\Sigma(p)\neq 0$ indicates that a dynamical mass is generated and chiral symmetry is broken.

Using Eq.~(\ref{SDfer}), we have
\begin{equation}\label{esdpf}
    S^{-1}_F(p)=-\slashed{p}+m-\Xi(p).
\end{equation}

Next, we replace Eq. \eqref{swdfe} into Eq. \eqref{esdpf} and obtain
\begin{equation}\label{mf}
    -A(p)\slashed{p}+\Sigma(p)=-\slashed{p}+m-\Xi(p).
\end{equation}
In order to calculate $A(p)$ and $\Sigma(p)$, we take the trace over the Dirac matrices \footnote{In (2+1)D Euclidean space and in the $4\times4$ representation: $\text{Tr}(\gamma^\mu,\gamma^\nu)=-4\delta^{\mu\nu}$, $\text{Tr}(\gamma^\mu,\gamma^\nu,\gamma^\alpha)=0$, and $\text{Tr}(\gamma^\mu,\gamma^\nu,\gamma^\alpha,\gamma^\beta)=4(\delta^{\mu\nu}\delta^{\alpha\beta}-\delta^{\mu\alpha}\delta^{\nu\beta}+\delta^{\mu\beta}\delta^{\nu\alpha})$.} \cite{Roberts1994}. After simple algebra, we find the mass function
\begin{equation}\label{fgmd}
    \Sigma(p)=m-\frac{\text{Tr}{\{\Xi(p)}\}}{\text{Tr}{\{\mathbb{I}}\}},
\end{equation}
where $\mathbb{I}$ is the identity matrix. To determine $A(p)$, we multiply Eq.~\eqref{swdfe} by $\slashed{p}$ and take again the trace over the Dirac matrices. Hence, we find
\begin{equation}\label{wfrf}
    A(p)=1+\frac{1}{\text{Tr}{\{\slashed{p}^2}\}}\text{Tr}{\{\slashed{p} \ \Xi (p)}\}.
\end{equation}

Eq.~(\ref{fgmd}) and Eq.~(\ref{wfrf}) are two coupled integral equations. The typical approximation used here is to consider $A(p)\rightarrow 1$, which is motivated by two reasons: first, it is a solution for the Ward-Takahashi identity, which relates both the vertex interaction $\Gamma^\nu(k,p)$ in the rainbow approximation and $S^{-1}_F(p)$, hence, it is in agreement with our previous assumptions \cite{Maris1996}. Furthermore, it also has a simple interpretation as long as we remember that $\Xi(p)\sim e^2$ and, therefore, we have
\begin{equation} \label{Aapro}
    A(p)=1+\mathcal{O}{\left(\frac{e^2}{4\pi}\right)}.
\end{equation}
Using Eq.~(\ref{Aapro}) in Eq.~(\ref{fgmd}) implies that we neglect higher-order contributions such as $\mathcal{O}(e^4)$ to $\Xi(p)$ when we take $A(p)=1$. Obviously, this is a perturbative approximation, but it is also in agreement with our previous simplifications. Furthermore, the nonperturbative feature remains well defined due to the arbitrariness of the mass function $\Sigma(p)$. Interestingly, from our numerical results, we also conclude that the second term on the right-hand side of Eq.~(\ref{wfrf}) becomes negligible in the large-external-momentum limit. In this case, it is enough to consider $A(p)=1$ (see \ref{B}) to calculate a nontrivial solution for $\Sigma(p)$. Similar results have been obtained for QED \cite{Roberts1994,Maris1995,Maris1996} and PQED \cite{Alves2013}, regarding the function $A(p)$. This approximation corresponds to the lowest-order solution of the coupled Schwinger–Dyson system and serves as a starting point for exploring the qualitative structure of the mass gap.

\section{Rainbow-Quenched Approximation}
\label{rainbowquenchedapproximation}

Within the rainbow-unquenched approximation we take $\Gamma^\mu\to-\gamma^\mu$, $\Delta_{\mu\nu}\to\Delta^0_{\mu\nu}$, and $A(p)\rightarrow 1$ \cite{Maris1996}. This set of approximations has recently been investigated in Ref.~\cite{Kzlers2015}, where it was shown to be in agreement with lattice calculations. Better vertex functions can lead to a different critical point, however the phase transition is always expected to occur. Therefore, from Eq.~(\ref{swdfe}), the full electron propagator reads
\begin{equation}
    S_F(p)=\frac{\slashed{p}+\Sigma(p)}{p^2+\Sigma^2(p)}.
\end{equation}
Furthermore, after some algebra and by using $m\to 0$, Eq. \eqref{fgmd} yields
\begin{align}\label{fgdmm}
    \Sigma(p)=4\pi\alpha\int\frac{d^{3}k}{(2\pi)^3}\frac{\delta^{\mu\nu}\Sigma(k)}{k^2+\Sigma^2(k)}\Delta^0_{\mu\nu}(p-k),
\end{align}
where we have defined $e^2=4\pi\alpha$. The constant $\alpha$ is known as the fine-structure constant and is a dimensionless fundamental constant that characterizes the strength of the electromagnetic interaction between two charged particles \cite{Alves2013}. In the context of the Schwinger-Dyson equations, it is relevant for studying dynamical mass generation, because we expect that such effect is realized only in the strongly correlated limit, i.e., for large values of $\alpha$ \cite{Roberts1994}.

Next, we use Eq. \eqref{pppqed} in Eq. \eqref{fgdmm}, within the Landau gauge ($\xi\to\infty$), hence the mass function reads
\begin{align}
    \Sigma(p)=4\pi\alpha&\int\frac{d^{3}k}{(2\pi)^3}\frac{\Sigma(k)}{k^2+\Sigma^2(k)}\times\nonumber\\
     &\times\left[\frac{1}{\sqrt{(p-k)^2}}-\frac{1}{\sqrt{(p-k)^2+\mu^2}}\right ].
\end{align}
Note that if $\mu\to0$, that is, $a\to\infty$, then the kernel of the integral vanishes, hence, $\Sigma(p)\rightarrow 0$, and no dynamical mass is generated. 

Thereafter, using spherical coordinates with $0\leq\varphi\leq 2\pi$, $0\leq\eta\leq \pi$, $0\leq k\leq \Lambda$, $d^{3}k=k^2\sin{\eta} dk d\varphi d\eta$,
and $(p-k)^2=p^2+k^2-2kp\cos{\eta}$,
we find
{\small \begin{align}
    &\Sigma(p)=\frac{4\pi\alpha}{(2\pi)^2}\int_0^\Lambda dk\frac{k^2\Sigma(k)}{k^ 2+\Sigma^2(k)}\int_0^\pi d\eta\frac{\sin{\eta}}{\sqrt{p ^2+k^2-2kp\cos{\eta}}}\nonumber\\
    &-\frac{4\pi\alpha}{(2\pi)^2}\int_0^\Lambda dk\frac{k^2\Sigma(k)}{k^2+\Sigma^ 2(k)}\int_0^\pi d\eta\frac{\sin{\eta}}{\sqrt{p^2+k^ 2-2kp\cos{\eta}+\mu^2}}.
\end{align}}
Note that we have included a UV cutoff $\Lambda$, which plays an important role for physical applications of our model, as we shall discuss later. After exactly solving the angular integral over $\eta$, we obtain
\begin{align}\label{fgdmmm}
    \Sigma(p)&=\frac{\alpha}{\pi p}\int_0^\Lambda dk\frac{k\Sigma(k)}{k^2+\Sigma^2( k)}K(p,k)+\nonumber\\
    &-\frac{\alpha}{\pi p}\int_0^\Lambda dk\frac{k\Sigma(k)}{k^2+\Sigma^2( k)}Q(p,k,\mu),
\end{align}
where we have defined $K(p,k)=|{k+p}|-|{k-p}|$ and $Q(p,k,\mu)=\sqrt{(k+p)^2+\mu^2}-\sqrt{(k-p)^2+\mu^ 2}$, which work as kernels to the integral equation for $\Sigma(p)$. From these definitions, it follows that
\begin{equation} \label{Kexp}
K(p,k)=\begin{cases}
    2p, \ k>p,\\
    2k, \ k<p,
\end{cases}
\end{equation}
and 
\begin{equation} \label{Mexp}
    Q(p,k,\mu)\approx \begin{cases}
    \frac{2pk}{(k^2+\mu^2)^{1/2}}, \ k\gg p,\\
    \frac{2pk}{(p^2+\mu^2)^{1/2}}, \ k\ll p.
\end{cases}
\end{equation}
In Eqs.~(\ref{Kexp}) and (\ref{Mexp}), we show the lowest-order contribution for each kernel as an expansion over both the external momentum $p$ and the loop momentum $k$. After using these results in Eq.~(\ref{fgdmmm}), we find an approximated integral equation for $\Sigma(p)$, namely,
\begin{align}\label{fsgdmi}
    \Sigma(p)&=\frac{2\alpha}{\pi p}\left[1-\frac{p}{(p^2+\mu^2)^{1/2}}\right]\int_0^{p} dk\frac{k^2\Sigma(k)}{k^2+\Sigma^ 2(k)}\nonumber\\
    &+\frac{2\alpha}{\pi}\int_{p}^{\Lambda} dk\frac{k\Sigma(k)}{k^2+\Sigma^2(k)}\left[1-\frac{k}{(k^2+\mu^2)^{1/2}}\right].
\end{align}
Obviously, this step is not mandatory for calculating a numerical solution, but it is relevant for an analytical calculation. Here, we should make a few comments about the parameter $\Lambda$. First, the term $\Lambda$ is a UV cutoff that regularizes the theory and allows us to manage the divergences that arise in loop integrals. Indeed, it is an upper limit for the acceptable energies within the physical system, hence, it can be interpreted as a scale beyond which our model is no longer a reasonable approximation \cite{Roberts1994}.

To obtain an analytical solution of Eq.~\eqref{fsgdmi}, we transform the integral equation into an equivalent differential equation with suitable asymptotic conditions for $\Sigma(p)$. Applying Leibniz's integral rule to  Eq.~(\ref{fsgdmi}), we can compute the first derivative of $\Sigma(p)$, given by
\begin{equation}\label{eqdfdm}
    \Sigma'(p)=-\frac{2\alpha}{\pi H(p,\mu) p^2}\int_0^{p} dk\frac{k^2\Sigma(k)}{k^2+\Sigma^2(k)},
\end{equation}
where we introduce the auxiliary functions
 $G(p,\mu)$ and $H(p,\mu)$ that shall be important. These are given by
\begin{equation}
    G(p,\mu)=\left[1-\frac{p}{(p^2+\mu^2)^{1/2}}\right]^{-1}
\end{equation}
and
\begin{equation}\label{funcH}
    H(p,\mu)=\left[1-\frac{p^3}{(p^2+\mu^2)^{3/2}}\right]^{-1}.
\end{equation}

The infrared boundary condition follows from the limit $p\to 0$ applied to Eq.~\eqref{eqdfdm} and find

\begin{equation}\label{ircond}
    \lim_{p\to0}\left[H(p,\mu) \ p^2\Sigma'(p)\right]=0.
\end{equation}
Similarly, the UV (ultraviolet) condition could be found by taking the limit $p\to\Lambda$ in Eq. \eqref{eqdfdm} and \eqref{fsgdmi}, after summing these equations. Therefore, we have
\begin{equation}\label{uvcond}
    \lim_{p\to\Lambda}\left[H(p,\mu) \ p^2\Sigma'(p)+G(p,\mu) \ p\Sigma(p)\right]=0.
\end{equation}

Differentiating Eq.~\eqref{eqdfdm} once more    and apply the fundamental theorem of calculus on the right-hand side, and after some algebra, we obtain
\begin{equation}\label{eqdfdmm}
    \frac{d}{dp}\left[H(p,\mu) \ p^2\Sigma'(p)\right]+\frac{2\alpha}{\pi}\frac{p^2\Sigma(p)}{p^2+\Sigma ^2(p)}=0,
\end{equation}
which is a differential equation for the mass function $\Sigma(p)$. This must also satisfy both the IR and UV conditions, given by Eqs. \eqref{ircond} and \eqref{uvcond}, respectively. Even after our set of approximations, Eq.~(\ref{eqdfdmm}) still is a nonlinear differential equation due to the denominator $p^2+\Sigma^2(p)$. This is a typical property of the Schwinger-Dyson equations, where the full propagator is written as a function of itself. Although Eq.~(\ref{eqdfdmm}) cannot be solved analytically in closed form, its asymptotic behavior can be investigated through physically motivated approximations. In order to proceed analytically, we now consider a sequence of controlled approximations.

\subsection{Approximations: $p\to0$}

The simplest approximation consists in assuming that the mass function $\Sigma(p)$ is a constant. From Eq.~(\ref{fsgdmi}), we conclude that this is a decreasing function of $p$, therefore, it is reasonable to consider the limit $p\to0$ in order to obtain an upper value for the dynamically generated mass. Due to this assumption, we take $\Sigma(p\to0)=M$ in Eq. \eqref{fsgdmi}, hence,
\begin{align}
    M=&\frac{2\alpha}{\pi}\int_{0}^{\Lambda} dk\frac{kM}{k^2+M^2}\nonumber+\\
    &-\frac{2\alpha }{\pi}\int_{0}^{\Lambda} dk\frac{k^2M}{k^2+M^2}\frac{1}{(k^2+\mu^ 2)^{1/2}},
\end{align}
which is our gap equation. After solving the integrals over $k$, we find
\begin{align}\label{spmc}
    \frac{\pi}{2\alpha}&=\frac{1}{2}\ln{\left(\frac{M^2+\Lambda^2}{M^2}\right)}-\ln{\left(\frac{\sqrt {\mu^2+\Lambda^2}+\Lambda}{\mu }\right)}+\nonumber\\
    &+\frac{|M|}{\sqrt{\mu^2-M^2}}\arctan{\left(\frac{\Lambda\sqrt{\mu^2-M^2}}{|M|\sqrt {\mu^2+\Lambda^2}}\right)}.
\end{align}
It is reasonable to assume that $\Lambda,\mu>>M$, which implies that the third term in the rhs of Eq. \eqref{spmc} is negligible. Therefore,
\begin{equation}
    \frac{\pi}{2\alpha}\approx \ln{\frac{\Lambda}{|M|}}-\ln{\left(\frac{\sqrt{\mu^2+\Lambda^2}+\Lambda}{ \mu}\right)},
\end{equation}
 hence,
\begin{equation}\label{spmcp}
    |M|\approx \frac{\Lambda\mu\exp\{-\frac{\pi}{2\alpha}\}}{\Lambda+\sqrt{\Lambda^2+\mu^2}}.
\end{equation}
The mass function given by Eq.~\eqref{spmcp}, vanishes when $\alpha\rightarrow 0$, as expected. Furthermore, it is cutoff independent in the limit $\Lambda\to\infty$, yielding a finite value $M\to(\mu/2)e^{-\pi/2\alpha}$. This indicates an improved ultraviolet behavior of $M$, when compared to the known results for PQED. Indeed, after taking $\mu\rightarrow\infty$ ($a\rightarrow 0$) in Eq.~(\ref{spmcp}), we find $M\to \Lambda e^{-\pi/2\alpha}$, which is a cutoff dependent result and represents the PQED result \cite{Alves2013}. 

\subsection{Approximation: $\Sigma^2(p)\gg p^2$}

We now consider the regime when $\Sigma(p)$ is not a constant. A simple approximation to the denominator of the second term in the lhs of Eq.~(\ref{eqdfdmm}) is to take $\Sigma^2(p)\gg p^2$. This linearizes the differential equation and allows us to obtain an analytical result. Hence, Eq. \eqref{eqdfdmm} reduces to
\begin{equation}
    \frac{d}{dp}\left[H(p,\mu) \ p^2\Sigma'(p)\right]\approx0
\end{equation}
whose solution is
\begin{align}
    \Sigma(p)&=C\left[-\frac{1}{p}+\frac{1}{\lambda}+\frac{1}{(p^2+\mu ^2)^{1/2}}-\frac{1}{(\lambda^2+\mu^2)^{1/2}}\right]+\nonumber\\
    &+\Sigma(\lambda),
\end{align}
where $C$ and $\lambda$ are arbitrary constants. Note that if $p=\lambda$, then $\Sigma(p)=\Sigma(\lambda)$. We obtain the constant $C$ after using the IR condition, given by Eq. \eqref{ircond}, which implies that the only acceptable value is $C=0$. Next, using the UV condition, given by Eq.~(\ref{uvcond}),  we conclude that $\Sigma(p)=\Sigma(\lambda)=0$. This means that we cannot describe dynamical mass generation in this approximation and only the trivial phase is a physical solution. The physical reason for such a result is that, for large external momentum, one would need $\Lambda\ll \Sigma(\Lambda)$, which is not reasonable since it also implies $\Lambda\ll M$, where $M$ is the maximum value of $\Sigma(p)$, as we have discussed in the previous approximation. Indeed, we expect that the cutoff is actually much larger than the value of the dynamical mass.

\subsection{Approximation: $\Sigma^2(p)\ll p^2$}

\begin{figure}
    \centering
    \includegraphics[width=\linewidth]{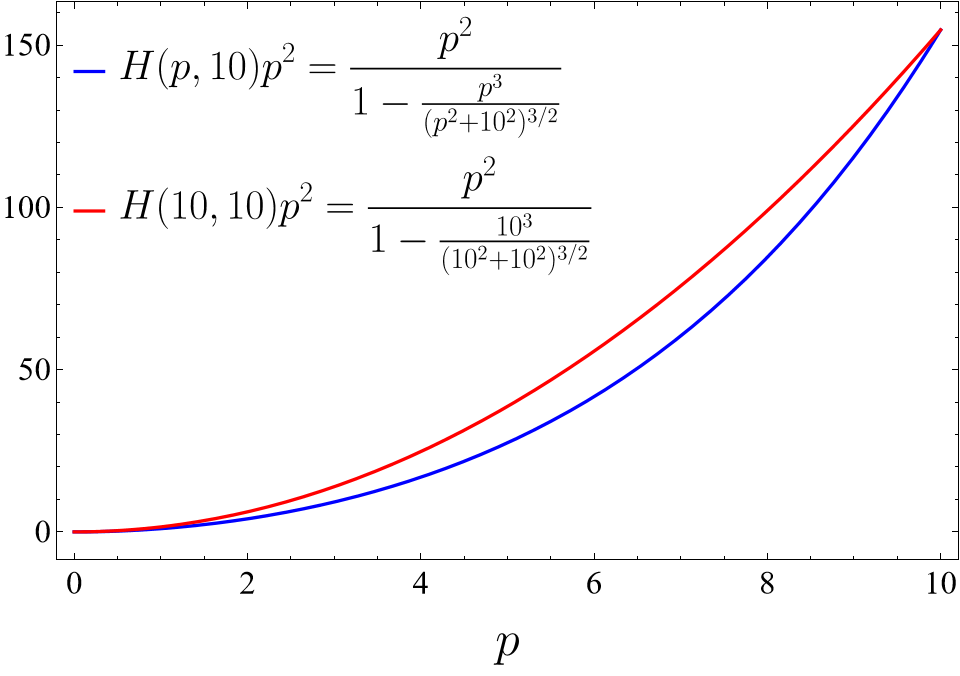}
    \caption{\textit{The $H(p,\mu)p^2\rightarrow H(\Lambda,\mu)p^2$ approximation within the large-momentum limit in Eq.~(\ref{EQDH})}. The blue line shows the exact expression, where we plot $p^2$ times Eq. \eqref{funcH} for $\mu=10$. The red line corresponds to the large-momentum approximation, plotting $p^2$ times Eq. \eqref{funcH} for $\mu=10$ and $\Lambda=10$. This approximation is valid as long as $\Lambda>p$ with a finite $\Lambda$.}
    \label{compgraphs}
\end{figure}

\begin{figure}
    \centering
    \includegraphics[width=\linewidth]{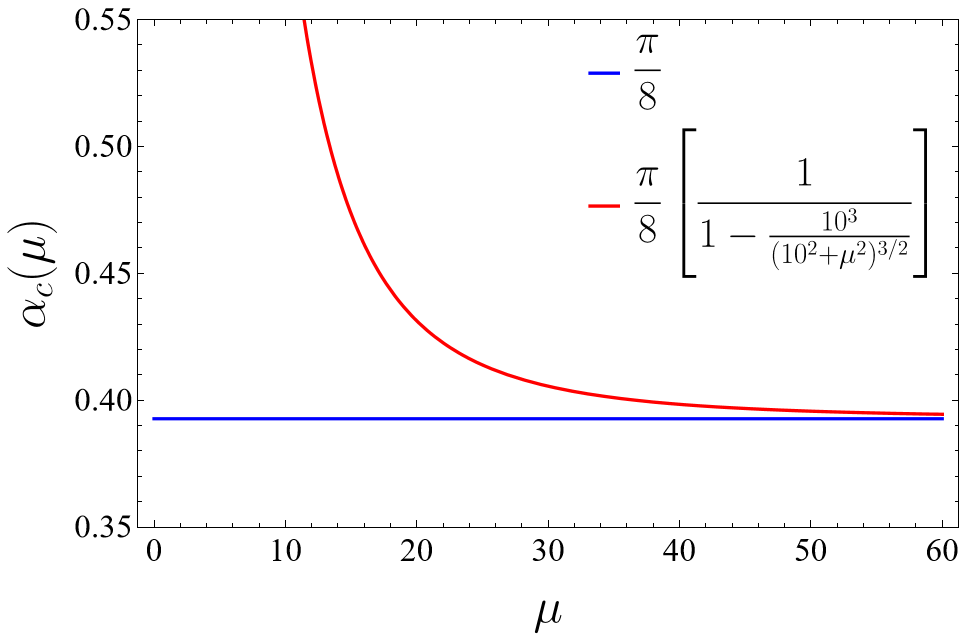}
    \caption{\textit{The critical coupling constant of PGQED in Eq.~(\ref{alcqr})}. The blue line shows Eq. \eqref{alcqr} with $\mu\rightarrow\infty$, which corresponds to the critical coupling constant in PQED \cite{Alves2013}. The red line shows Eq. \eqref{alcqr} for $\Lambda=10$. We observe that as $\mu$ decreases, $\alpha_c$ increases making dynamical mass generation increasingly difficult to achieve because it needs $\alpha>\alpha_c(\mu)$.}
    \label{alphacrit}
\end{figure}

Motivated by the previous analysis, the most realistic approximation to the term $p^2+\Sigma^2(p)$ would be $p^2+\Sigma^2(p)\approx p^2+M^2$. This works well for both the small- and large-momentum limits. For analytical simplicity, we restrict ourselves to the limiting case where $\Sigma^2(p)\ll p^2$, which is sufficient to describe the mass function, as we shall show in this section. We also consider a simplification $H(p,\mu) \ p^2\to H(\Lambda,\mu) \ p^2$, which is a good approximation as long as $\Lambda>p$, see Fig.~\ref{compgraphs}. Under these approximations, from Eq. \eqref{eqdfdmm}, we obtain a Cauchy-Euler differential equation, namely,
\begin{equation} \label{EQDH}
    p^2\Sigma''(p)+2p\Sigma'(p)+\frac{2\alpha}{\pi H(\Lambda,\mu)}\Sigma(p)=0.
\end{equation}
The solution of this equation reads

\begin{equation}\label{fgdmee}
    \Sigma(p)=A_1p^{\gamma_{-}}+A_2p^{\gamma_{+}},
\end{equation}
where $(A_1,A_2)$ are arbitrary constants,
\begin{equation}
    \gamma_\pm=-\frac{1}{2}\pm\frac{1}{2}\sqrt{1-\frac{\alpha}{\alpha_c}}=-\frac{1}{2}\pm i\beta,
\end{equation}
$\beta=\frac{1}{2}\sqrt{\frac{\alpha}{\alpha_c}-1}$, and
\begin{equation} \label{alcqr}
    \alpha_c=\frac{\pi H(\Lambda,\mu)}{8}=\frac{\pi}{8}\left[\frac{1}{1-\frac{\Lambda^3}{( \Lambda^2+\mu^2)^{3/2}}}\right].
\end{equation}
The constant $\alpha_c$, identified as the critical coupling constant, defines the threshold above which the mass function becomes oscillatory. This regime $\alpha>\alpha_c$, corresponds to the strong-coupling phase where dynamical chiral symmetry breaking is expected to occur. The Podolsky parameter plays an important role in determining this threshold. Physically, the increase of $\alpha_c$ with decreasing $\mu$ reflects the screening induced by the Podolsky sector, which weakens the effective long-range interaction responsible for dynamical mass generation. Consequently, stronger coupling is required to overcome this screening and generate a nonzero fermion mass. Fig.~\ref{alphacrit} illustrates this behavior: as $\mu\rightarrow\infty$, one recovers the PQED result $\alpha_c=\pi/8$~\cite{Alves2013}, whereas decreasing $\mu$ leads to a monotonic increase of the critical coupling, making dynamical mass generation progressively more difficult.

For $\alpha>\alpha_c$, the mass function can be written, after some algebra, as
\begin{equation} \label{Sig0}
    \Sigma(p)=\frac{ C}{\sqrt{p}}\sin{[\beta(\ln{\frac{p}{\Sigma_0}}+\delta)]},
\end{equation}
where $B_1=A_1+A_2$, $B_2=A_1-A_2$, $C=\sqrt{B_1^2-B_2^2}$, and $\delta=\arctan{(-iB_1/B_2)}$. The scale factor $\Sigma_0$ is introduced due to the logarithmic term. The scale is determined from the ultraviolet boundary condition, given by Eq. \eqref{uvcond}, after considering the same approximations made to the differential equation. Hence, we have
\begin{equation}\label{uvcond2}
    \lim_{p\to\Lambda}\left[H(\Lambda,\mu) \ p^2\Sigma'(p)+G(\Lambda,\mu) \ p\Sigma(p)\right]=0.
\end{equation}
After using Eq.~(\ref{Sig0}) in Eq.~(\ref{uvcond2}), we obtain
\begin{align} \label{Miran}
    \Sigma_0&=\Lambda\exp\left\{\delta+\frac{2 \left(\Lambda^2+\mu^2\right)}{2\Lambda\sqrt{\Lambda^2+\mu^2} +3 \Lambda ^2+\mu ^2}\right\}\times\nonumber\\
    &\times\exp\left\{\frac{-2n\pi}{\sqrt{\frac{\alpha}{\alpha_c}-1}}\right\},
\end{align}
where $n=0,1,2,...$ is an integer number. The scale factor in Eq.~(\ref{Miran}) obeys a Miransky scaling law, whose main feature is that $\Sigma_0$ remains finite in the limit $\Lambda\rightarrow\infty$ as $\alpha\rightarrow \alpha_c$. This essential singularity at the critical point is one of the characteristic signatures of dynamical symmetry breaking in strongly coupled gauge theories. For $\alpha<\alpha_c$, it is straightforward to show that the mass function does not obey the UV condition, hence, the system must remain in the massless phase. This is in agreement with previous results for both QED and PQED. Since $\alpha_c(\mu)$ is always larger than the PQED value $\alpha_c(\infty)=\pi/8$, it follows that the Podolsky parameter does not favor dynamical mass generation.

\section{Rainbow-Unquenched Approximation}
\label{rainbowunquenchedapproximation}

As we have discussed before, within this approximation, we take $\Gamma^\mu\to-\gamma^\mu$, $\Delta_{\mu\nu}$ at lowest order in the large-$N$ expansion, and $A(p)\rightarrow 1$ \cite{Maris1996}. Therefore, the main modification is to consider a corrected gauge-field propagator \cite{Montenegro2020,Appelquist1985}.

We begin by deriving how the vacuum polarization tensor modifies the gauge-field propagator. Within the four-rank representation of Dirac matrices, the vacuum polarization tensor reads
\begin{equation}
    \Pi^{\mu\nu}(p,m)=\Pi_1(p,m)P^{\mu\nu},
\end{equation}
where the projection operator is given by
\begin{equation}
    P_{\mu\nu}=\delta_{\mu\nu}-\frac{p_\mu p_\nu}{p^2}.
\end{equation}
Using the Schwinger-Dyson equation for the gauge-field propagator in Eq. \eqref{esdpp}, we obtain
\begin{equation}
    \Delta_{\mu\nu}=\Delta_{0\mu\alpha}\{G^{\alpha}_{\nu}\}^{-1},
\end{equation}
where
\begin{equation} \label{1N}
    G^{\alpha}_{\nu}=\delta_{\nu}^{\alpha}-\Pi^{\alpha\beta}\Delta_{0\beta\nu}=\delta_{\nu}^ {\alpha}-\Pi_1 P^{\alpha\beta}\Delta_{0\beta\nu}.
\end{equation}
After some simplifications, we find
\begin{equation}\label{PropUnq}
    \Delta_{\mu\nu}=\frac{P_{\mu\nu}}{\left[\frac{1}{2}\frac{1}{(p^2)^{1/2}}-\frac{1}{2}\frac{1}{(p ^2+\mu^2)^{1/2}}\right]^{-1}-\Pi_1}.
\end{equation}
For massless fermions $m\to0$, the function $\Pi_1(p)$ is \cite{Kondo1995}
\begin{equation}
    \Pi_1(p)=-\frac{g}{8}(p^2)^{1/2},
\end{equation}
where we define $g\rightarrow e^2 N$ as a coupling constant related to the electric charge $e$ and to the flavor number of the Dirac field $N$. Note that we could also consider $g\rightarrow 4\pi\alpha N$ written in terms of the fine-structure constant. The main idea behind the large-$N$ expansion is that $g$ remains fixed when $\alpha\ll 1$ and $N\gg 1$ \cite{Coleman1985}. 

Therefore, the corrected gauge-field propagator is given by
\begin{equation}\label{pcppqed}
    \Delta_{\mu\nu}=\frac{P_{\mu\nu}}{\left[\frac{1}{2}\frac{1}{(p^2)^{1/2}} -\frac{1}{2}\frac{1}{(p^2+\mu^2)^{1/2}}\right]^{-1}+\frac{g}{8}( p^2)^{1/2}},
\end{equation}
where we have used the Landau gauge $\xi\to\infty$. Using these approximations and Eq. \eqref{fgmd}, we find that the mass function is given by
\begin{equation}\label{fgdmmc}
    \Sigma(p)=\frac{g}{N}\int\frac{d^{3}k}{(2\pi)^3}\frac{\Sigma(k)}{k^2+\Sigma^2(k)}\delta^{\mu\nu}\Delta_{\mu\nu}(p-k).
\end{equation}
Using Eq. \eqref{pcppqed} in Eq. \eqref{fgdmm}, and keeping in mind that $4\pi\alpha\rightarrow g/N$, we find
\begin{align} \label{aux1}
    \Sigma(p)&=\frac{2g}{N}\int\frac{d^{3}k}{(2\pi)^3}\frac{\Sigma(k)}{k^2+\Sigma^2(k)}\times\nonumber\\
    &\times\frac{1}{\left[\frac{1}{2}\frac{1}{\sqrt{(p-k)^2}}-\frac{1}{2}\frac{1}{\sqrt{(p-k)^2+\mu^2}}\right]^{-1}+\frac{g}{8}{\sqrt{(p-k)^2}}}.
\end{align}
As in the rainbow-quenched approximation, we conclude that when $\mu\to0$ ($a\to\infty$), the kernel of the integral in Eq.~(\ref{aux1}) vanishes. Using spherical coordinates for the loop integral, we have
\begin{equation} \label{SigIntN}
    \Sigma(p)=\frac{4\pi g}{(2\pi)^3N}\int_0^\Lambda \int_0^\pi dkd\eta\frac{k^2\Sigma(k)}{k^2+\Sigma^2(k)} \ L(p-k,\eta,\mu,g),
\end{equation}
where
\begin{equation}
    L(q,\eta,\mu,g)=\frac{\sin{\eta}}{\left[\frac{1}{2}\frac{1}{\sqrt{q^2}}-\frac{1}{2}\frac{1}{\sqrt{q^2+\mu^2}}\right]^{-1}+\frac{g}{8}{\sqrt{q^2}}}
\end{equation}
is the kernel of the integral and $q=p-k$. Next, let us solve the integral equation for $\Sigma(p)$ using the same momentum expansion applied before.

\subsection{Approximation: $\Sigma^2(p)\ll p^2$}

To obtain an analytical treatment analogous to the rainbow-quenched case, we expand the kernel in the asymptotic momentum regimes $k\gg p$ and $k\ll p$. Hence, for $k\gg p$, we have
\begin{equation}
    L(k,\eta,\mu,g)\approx \frac{\sin\eta}{\frac{g \sqrt{k^2}}{8} +\frac{1}{\frac{1}{2 \sqrt{k^2}}-\frac{1}{2 \sqrt{k^2+\mu ^2}}}},
\end{equation}
while for $k\ll p$ it reads
\begin{equation}
    L(p,\eta,\mu,g)\approx \frac{\sin\eta}{\frac{g \sqrt{p^2}}{8} +\frac{1}{\frac{1}{2 \sqrt{p^2}}-\frac{1}{2 \sqrt{p^2+\mu^2}}}}.
\end{equation}

This expansion allows angular integration to be carried out analytically, while preserving the dominant asymptotic behavior of the kernel. Therefore, we have
\begin{align}
    \Sigma(p)&=\frac{4\pi g}{(2\pi)^3N}\int_0^\Lambda\int_0^\pi dk d\eta\frac{k^2 \Sigma(k)}{k^2+\Sigma^2(k)}\times\nonumber\\
    &\times\Bigg\{L(k,\eta,\mu,g)\theta(k-p)+L(p,\eta,\mu,g)\theta(p-k)\Bigg\},
\end{align}
where $\theta(x)$ is the Heaviside step function. After solving the angular integral, we obtain
\begin{align}\label{eifgdmrq}
    \Sigma(p)&=\frac{8\pi g}{(2\pi)^3N}\int_0^p dk\frac{k^2\Sigma(k)}{k^2+\Sigma^2(k)}\left\{\frac{I(p,\mu,g)}{p}\right\}\nonumber\\
    &+\frac{8\pi g}{(2\pi)^3N}\int_p^\Lambda dk\frac{k^2\Sigma(k)}{k ^2+\Sigma^2(k)}\left\{\frac{I(k,\mu,g)}{k}\right\}.
\end{align}
where the auxiliary function $I(p,\mu,g)$ is defined as
\begin{equation}
    I(p,\mu,g)=\left[\frac{g}{8}+\frac{1} {\frac{1}{2}-\frac{p}{2\sqrt{p^2+\mu^2}}}\right]^{-1}.
\end{equation}
Differentiating Eq.~\eqref{eifgdmrq} with respect to $p$, it follows that
\begin{equation}\label{pdfgdmrq}
    \frac{J(p,\mu,g)}{I^{2}(p,\mu,g)}p^2\Sigma' (p)
    +\frac{8\pi g}{(2\pi)^3N}\int_0^p  dk \frac{k^2 \Sigma(k)}{k^2+\Sigma ^2(k)}=0,
\end{equation}
with
\begin{equation}
    J(p,\mu,g)=\left[\frac{g}{8}-\frac{\frac{p}{2 \left(p^2+\mu^2\right)^{3/2}}-\frac{1}{2p^2}}{\left(\frac{1}{2p}-\frac{1}{2 \sqrt{p^2+\mu^2}}\right)^2}\right]^{-1},
\end{equation}
which plays a role analogous to the function $H(p,\mu)$ introduced in the rainbow-quenched approximation. Differentiating once more with respect to $p$ yields
\begin{align}\label{edfgdmrq}
    \frac{d}{dp}\left[\frac{J(p,\mu,g)}{I^{2}(p,\mu,g)}p^2\Sigma'(p)\right]+\frac{8\pi g}{(2\pi)^3N} \frac{p^2\Sigma(p)}{p^2+\Sigma^2(p)}=0.
\end{align}
As in the rainbow-quenched approximation, we consider $\Sigma^2(p)\ll p^2$ and 
\begin{equation}
    \frac{J(p,\mu,g)}{I^{2}(p,\mu,g)}p^2\to \frac{J(\Lambda,\mu,g)}{I^{2}(\Lambda,\mu,g)}p^2
\end{equation}
in order to obtain a Cauchy–Euler differential equation, given by
\begin{equation} \label{EqdifN}
    p^2\Sigma''(p)+2p\Sigma'(p)+\frac{I^2(\Lambda,\mu,g)}{J(\Lambda,\mu,g)}\frac{g}{\pi^2 N }\Sigma(p)= 0.
\end{equation}
The resulting differential equation has the same Cauchy-Euler structure obtained in the rainbow-quenched approximation, with the Podolsky and vacuum-polarization effects encoded in the effective coefficient $I^2/J$. The solution of Eq.~(\ref{EqdifN}) reads
\begin{equation}\label{fgdmeee}
    \Sigma(p)=E_1p^{\gamma_{-}}+E_2p^{\gamma_{+}},
\end{equation}
where $(E_1,E_2)$ are arbitrary constants,
\begin{equation}
    \gamma_\pm=-\frac{1 }{2}\pm\frac{1}{2}\sqrt{1-\frac{N_c}{N}}=-\frac{1}{2}\pm i\beta,
\end{equation}
with $\beta=\frac{1}{2}\sqrt{\frac{N_c}{N}-1}$, and
\begin{equation} \label{NcN}
    N_c=\frac{4g}{\pi^2}\frac{I^2(\Lambda,\mu,g)}{J(\Lambda,\mu,g)}=\frac{4g}{\pi^2}\frac{\left[g/8+\frac{y_{1/2}^2(\Lambda,\mu)}{y_{3/2}(\Lambda,\mu)}\right]}{\left[g/8+y_{1/2}(\Lambda,\mu)\right]^2},
\end{equation}
is the critical number of fermion flavors, below which the theory is expected to develop a dynamically generated fermion mass, and
\begin{equation}
    y_{x}(\Lambda,\mu)=\left[\frac{1}{2}-\frac{1}{2}\left(\frac{\Lambda^2}{\Lambda^2+\mu^2}\right)^x\right]^{-1},
\end{equation}
is an auxiliary function.
\begin{figure}
    \centering
    \includegraphics[width=\linewidth]{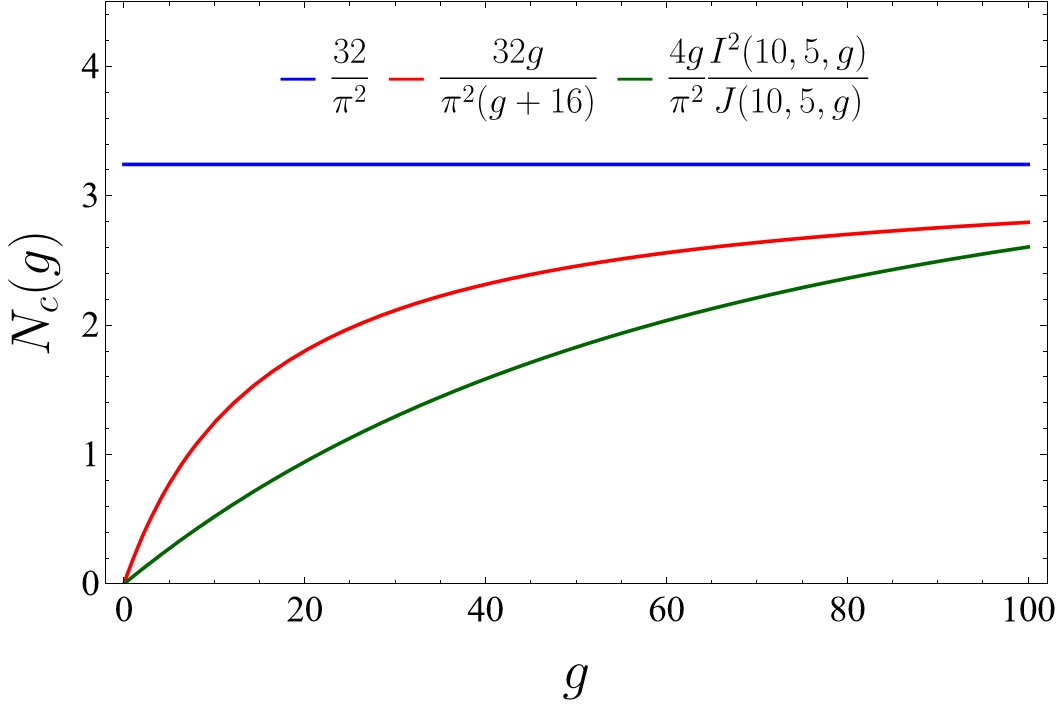}
    \caption{\textit{The critical coupling constant of PGQED in Eq.~(\ref{NcN})}. The blue line shows Eq. \eqref{NcN} with $\mu\rightarrow\infty$ and $g\rightarrow\infty$, which corresponds to the critical coupling constant in QED3 \cite{Appelquist1985}. The red line shows Eq. \eqref{NcN} with $\mu\rightarrow\infty$, which corresponds to the critical coupling constant in PQED \cite{Alves2013}. The green line shows Eq. \eqref{NcN} for $\Lambda=10$ and $\mu=5$. We observe that the function has an upper limit and that $N_c$ increases as $g$ increases.}
    \label{ncrit}
\end{figure}

For $\mu \gg\Lambda$ ($a\Lambda \ll 1$), we have $N_c\rightarrow \frac{32g}{\pi^2(g+16)}$, which is the same result as that obtained in PQED within the rainbow-unquenched approximation \cite{Alves2013}. We can also see this behavior in Fig.~\ref{ncrit}. In the strong-coupling limit $g\gg 1$, we have $N_c\rightarrow 32/\pi^2$, which is the critical number for QED$_3$ \cite{Appelquist1985}. From Eq.~(\ref{NcN}), we may also conclude that $N_c\rightarrow 0$ when $\mu\ll \Lambda$ ($a\Lambda \gg 1$), hence, no dynamical mass generation occurs. Indeed, for $N>N_c$ there is no chiral symmetry breaking, which is a consequence of the boundary conditions, given by
\begin{equation}
    \lim_{p\to\Lambda}\left[J(p,\mu,g) \ p\Sigma'(p)+I(p,\mu,g)\Sigma(p)\right]=0
\end{equation}
in the UV limit, and
\begin{equation}\label{uvcondm}
    \lim_{p\to0}\left[\frac{J(p,\mu,g)}{I^{2}(p,\mu,g)} \ p^2\Sigma'(p)\right]=0
\end{equation}
in the IR regime. Following the same steps as before, we also obtain a scale parameter $\bar\Sigma_0$, given by
{\scriptsize \begin{align} \label{scale}
    \bar\Sigma_0&=\Lambda\exp\left\{\bar\delta+\frac{2 \left(-g \Lambda\sqrt{\Lambda^2+\mu^2}+(g+16) \Lambda^2+(g +16) \mu ^2\right)}{-(g-32) \Lambda \sqrt{\Lambda ^2+\mu ^2}+(g+48) \Lambda ^2+(g+16) \mu ^2}\right\}\times\nonumber\\
    &\times\exp\left\{\frac{-2n\pi}{\sqrt{\frac{N_c}{N}-1 }}\right\},
\end{align}} 
where $\bar\delta$ is a function of $(E_1,E_2)$ and $n=0,1,2...$ is an integer number. The scale in Eq.~(\ref{scale}) is real and finite only if $N<N_c$ and has been calculated in the limit $N\approx N_c$ ($\beta\approx 0$).

Next, let us discuss one possible explanation why our $N_c$ agrees with QED3 at large interactions. This has also been observed in PQED, and it seems to indicate that, within this asymptotically-large-coupling limit, the electron-electron interaction is actually dominated by the vacuum polarization tensor for a massless electron. Indeed, the corrected gauge-field propagator for a gauge theory with a U(1) interaction generally takes the form $\propto 1/(G(p^2)+g p)$ in the large-$N$ expansion, where $G(p^2)$ is the momentum function of the bare gauge-field propagator of these kinds of models in (2+1)D, for example, QED ($G\propto p^2$), PQED ($G\propto \sqrt{p^2}$), and PGQED (see Eq.~(\ref{PropUnq})).
On the other hand, the term $gp$ is the contribution from the vacuum polarization tensor for massless fermions, which does not depend on the gauge-field propagator at one-loop approximation. Hence, whenever $gp\gg G(p^2)$ ($g \rightarrow \infty$), all of these gauge-field propagators would behave similarly and are expected to generate the same critical number (or a very similar value).

\section{Ball-Chiu vertex approximation}
\label{ballchiuvertexapproximation}

Up to this point, we have employed the bare interaction vertex within the rainbow-ladder approximation. However, this choice violates the Ward-Takahashi identity (WTI). To assess the robustness of the rainbow approximation of our approximation, we also incorporate the Ball-Chiu (BC) vertex. Unlike the bare vertex, the BC construction is designed to satisfy the WTI identically, thereby preserving the fundamental gauge invariance of the theory at the non-perturbative level. This vertex is defined as:
\begin{equation}
    \Gamma_\mu^{BC}(p,k) = f_0\gamma_\mu + f_1(\slashed{p}+\slashed{k})(p+k)_\mu + f_2(p+k)_\mu,
\end{equation}
where the scalar functions $f_0$, $f_1$, and $f_2$ are given by
\begin{equation}
    f_0 =- \frac{A(p)+A(k)}{2},
\end{equation}
\begin{equation}
    f_1 =- \frac{A(p)-A(k)}{2(p^2-k^2)},
\end{equation}
\begin{equation}
    f_2 = \frac{\Sigma(p)-\Sigma(k)}{p^2-k^2}.
\end{equation}

Working in the Landau gauge and neglecting the effects of the vacuum polarization tensor (the quenched approximation), the coupled integral equations for the mass gap $\Sigma(p)$ and the wave-function renormalization $A(p)$ are
\begin{align}
    \Sigma(p) &= \frac{\alpha}{\pi}\int_0^\Lambda\frac{dk\,k^2}{A^2(k)k^2+\Sigma^2(k)}\times\nonumber\\
    &\times\left(- \Sigma(k)f_0 I_1 - p^2 k^2 (2\Sigma(k)f_1 + A(k)f_2) I_2 \right)
\end{align}
and
\begin{align}
    A(p) &= 1 + \frac{\alpha}{\pi p^2}\int_0^\Lambda\frac{dk\,k^2}{A^2(k)k^2+\Sigma^2(k)}\times\nonumber\\
    &\times( -A(k)f_0 p k I_3 + \nonumber\\
    &+k^2 p^2 (A(k)f_0 + A(k)f_1(p^2+k^2) - \Sigma(k)f_2) I_2),
\end{align}
where the angular integrals are defined by
\begin{align}\label{angint1}
    I_1 &= \int_0^{\pi} d\theta \sin\theta \left[ \frac{1}{\sqrt{r(p,k,\theta)}} - \frac{1}{\sqrt{r(p,k,\theta)+\mu^2}} \right],
\end{align}
\begin{align}\label{angint2}
    I_2 &= \int_0^{\pi} d\theta \frac{\sin\theta(1-\cos^2\theta)}{r(p,k,\theta)}\times\nonumber\\
    &\times\left[ \frac{1}{\sqrt{r(p,k,\theta)}} - \frac{1}{\sqrt{r(p,k,\theta)+\mu^2}} \right],
\end{align}
\begin{align}\label{angint3}
    I_3 &= \int_0^{\pi} d\theta \sin\theta\cos\theta \left[ \frac{1}{\sqrt{r(p,k,\theta)}} - \frac{1}{\sqrt{r(p,k,\theta)+\mu^2}} \right].
\end{align}
where $r(p,k,\theta)=p^2+k^2-2pk\cos\theta$. Unlike the rainbow-ladder approximation, where $A(p)=1$ is assumed, this framework requires solving the coupled equations for both $\Sigma(p)$ and $A(p)$ numerically.

Now, when incorporating the effects of the vacuum polarization tensor (evaluated at leading order in the $1/N$ expansion), the system of equations becomes
\begin{align}
    \Sigma(p) &= \frac{g}{2\pi^2 N}\int_0^\Lambda\frac{dk\,k^2}{A^2(k)k^2+\Sigma^2(k)}\times\nonumber\\
    &\times\left(-\Sigma(k)f_0 I_1^\ast - p^2 k^2 (2\Sigma(k)f_1 + A(k)f_2) I_2^\ast \right)
\end{align}
and
\begin{align}
    A(p) &= 1 + \frac{g}{2\pi^2 N p^2}\int_0^\Lambda\frac{dk\,k^2}{A^2(k)k^2+\Sigma^2(k)}\times\nonumber\\
    &\times( -A(k)f_0 p k I_3^\ast + \nonumber\\
    & + k^2 p^2 (A(k)f_0 + A(k)f_1(p^2+k^2) - \Sigma(k)f_2) I_2^\ast ).
\end{align}

The modified angular integrals $I_1^\ast$, $I_2^\ast$, and $I_3^\ast$ share a similar structure to Eqs.~\eqref{angint1}, \eqref{angint2}, and \eqref{angint3}, respectively, but with the interaction kernel replaced by the full effective propagator:
\begin{equation}
    \left[ \left( \frac{1}{2\sqrt{r(p,k,\theta)}} - \frac{1}{2\sqrt{r(p,k,\theta)+\mu^2}} \right)^{-1} + \frac{g}{8}\sqrt{r(p,k,\theta)} \right]^{-1}.
\end{equation}
In the following subsection, we perform a numerical analysis of the scalar functions $\Sigma(p)$ and $A(p)$.

\subsection{Numerical results}

The inclusion of the Ball-Chiu vertex significantly increases the mathematical complexity of the Schwinger-Dyson equations. Unlike the bare vertex approximation, the coupled nature of the mass gap $\Sigma(p)$ and the wave-function renormalization $A(p)$, combined with the highly non-linear structures emerging from the Podolsky-regularized angular integrals, renders an analytical solution in closed form intractable. Consequently, to accurately evaluate the dynamical chiral symmetry breaking and capture the subtle competition between the radiative enhancement of the vertex and the shielding effect of the Podolsky mass $\mu$, we must resort to robust numerical techniques.

To solve this intricate non-linear system, we employ a numerical framework based on momentum space discretization. The momentum variables are defined over an integration domain bounded by an infrared scale $p_{\rm{min}} = 10^{-3}$ and an ultraviolet cutoff $\Lambda = 10$. By discretizing this interval into a grid of $M=300$ momentum points and applying the repeated trapezoidal quadrature rule, the coupled integral equations are systematically converted into a large set of algebraic equations. This discretized system is then solved numerically to simultaneously yield the momentum-dependent profiles of both $\Sigma(p)$ and $A(p)$ for a given coupling strength. Similar numerical procedures have been successfully applied in previous investigations of dynamical mass generation in PQED \cite{Alves2013, Nascimento2015}. Further details are provided in~\ref{B}.

\begin{figure}
    \centering
    \includegraphics[width=\linewidth]{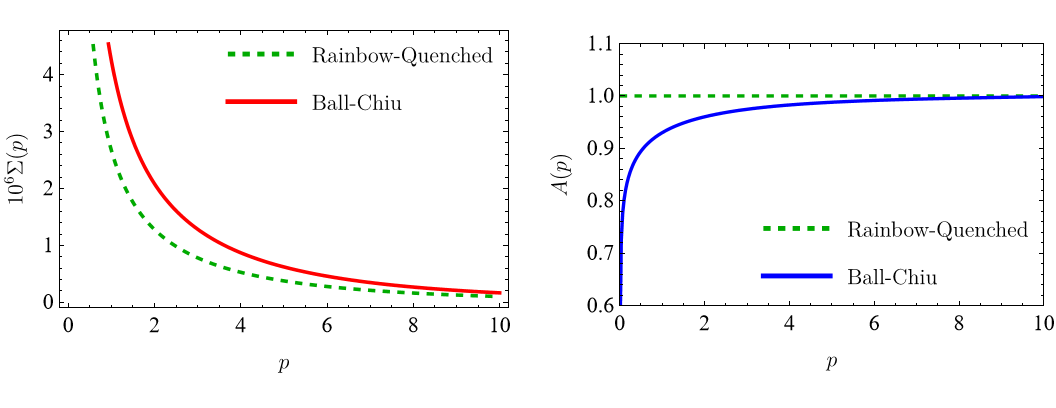}
    \caption{\textit{Numerical solutions for the mass gap $\Sigma(p)$ and wave-function renormalization $A(p)$ in PGQED}. The left panel shows $\Sigma(p)$ as a function of $p$ for the rainbow-quenched (dashed green) and Ball-Chiu (solid red) approximations. The right panel shows $A(p)$, which is strictly $1$ for the quenched case (dashed green) but dynamically enhanced in the infrared by the BC vertex (solid blue). Parameters are set to $\alpha = 0.25$, $\mu = 10$, and $\Lambda = 10$.}
    \label{ppqedbcvsquenched}
\end{figure}

\begin{figure}
    \centering
    \includegraphics[width=\linewidth]{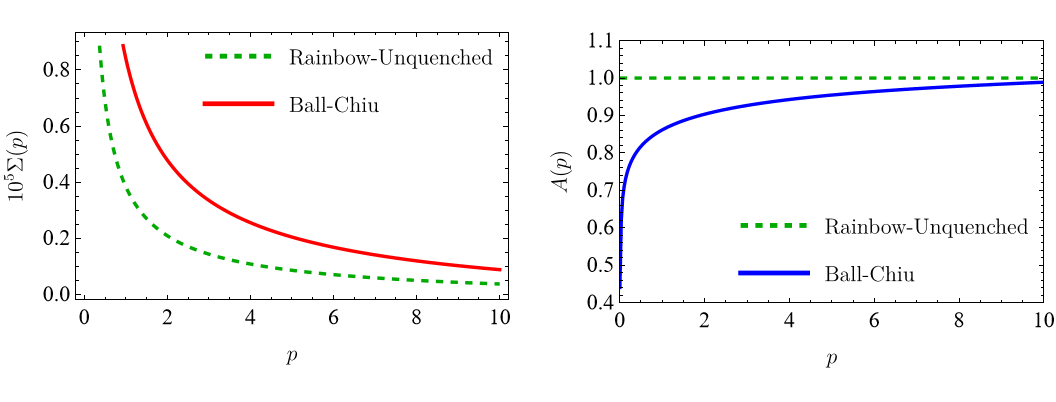}
    \caption{\textit{Numerical solutions for the mass gap $\Sigma(p)$ and wave-function renormalization $A(p)$ in unquenched PGQED}. On the left panel, we have $\Sigma(p)$ as a function of $p$ for the rainbow-unquenched (dashed green) and unquenched Ball-Chiu (solid red) approximations. On the right panel, we have $A(p)$, which is fixed at $1$ for the rainbow-unquenched case (dashed green) but dynamically dressed by the BC vertex (solid blue). Parameters are set to $g=100$, $N_f=4$, $\mu=10$, and $\Lambda=10$.}
    \label{ppqedbcvsunquenched}
\end{figure}


As evidenced by the numerical results shown in Figs. \ref{ppqedbcvsquenched} and \ref{ppqedbcvsunquenched}, the inclusion of the Ball-Chiu vertex introduces significant quantitative modifications to the dynamical mass generation process, notably affecting the absolute values of the gap function $\Sigma(p)$. However, it is crucial to emphasize that this sophisticated vertex construction does not modify the qualitative structure of the solutions. The asymptotic behavior and the overall momentum dependence of $\Sigma(p)$ remain structurally identical to those observed under the bare vertex approximation. Therefore, although the non-trivial dressing of the wave-function renormalization $A(p)$ provides a more complete and gauge-invariant framework, the fundamental physical picture of chiral symmetry breaking in PGQED remains qualitatively robust against these higher-order vertex corrections.

\section{Final remarks}
\label{finalremarks}

We have shown that chiral symmetry breaking occurs in PGQED in the strong-coupling limit, which is obtained when the coupling constant is larger than a critical value, i.e., $\alpha>\alpha_c(\mu)=H(\Lambda,\mu)\pi/8$, where $H(\mu,\Lambda)\geq 1$ is a known function within the rainbow-quenched approximation. Similar results have been derived when quantum corrections are included in the gauge-field propagator, using the large-$N$ expansion. In this case, with $g\rightarrow \infty$ (strong interactions), we have obtained a critical point $N_c=32/\pi^2$, which is exactly the same critical point obtained earlier for QED3 \cite{Appelquist1985,Roberts1994,Maris1996}. Thereafter, we have used the Ball-Chiu-vertex approximation in order to calculate the mass function. From our numerical results, we have concluded that the chiral phase transition is qualitatively well described by the bare-vertex approximations. Motivated by possible applications to graphene, we have included a Fermi velocity $v_F$ in our model in \ref{C}. In this case, we have shown that the critical coupling constant increases in comparison to the isotropic case. One important extension of our results would be the inclusion of an external magnetic field in order to describe its effects into the chiral phase transition in both PGQED and two-dimensional materials \cite{Rojas2008,Menezes2016}. We shall discuss this elsewhere. 


Next, let us discuss some important results related to PGQED in literature. Recently, Ref. \cite{Neves2025} performed a dimensional reduction of Lee-Wick electrodynamics from (3+1)D to (2+1)D. This calculation yields the same reduced gauge-field propagator we have obtained here, given by Eq.~(\ref{pppqed}) in momentum space, although the starting action in (3+1)D is slightly different from ours. The two formulations are physically equivalent at the level of the static interaction, since both lead to the same potential $V(r)\propto (1-e^{-\mu r})/r$. In Ref. \cite{Neves2025}, it is also shown that this kind of model respects the microcausality at tree level, which is a desirable feature for a well-defined field theory. Here, our main results depend on the strength of quantum corrections and are likely to be the same in the context of the reduced Lee-Wick electrodynamics, given the similarity of its gauge-field propagator in comparison to the one in PGQED.

The emergence of gauge theories in low dimensions has also been largely discussed in the context of quantum simulations with ultracold atoms. Interestingly, a proposal for the realization of QED in two spatial dimensions, using a mixture of ultracold atoms trapped in an optical lattice, has been discussed in Ref.~\cite{Ott2021}. Possible experimental setups and lattice gauge theories have been discussed for quantum computation in Refs.~\cite{Halimeh2025,Meth2025}. Our action in PGQED is a useful example of an emerging gauge theory as it is defined by a pseudo-differential operator. This, in turn, yields a richer form of static interaction among electrons and further generalizations are straightforward. Furthermore, we have two main parameters $\Lambda$ and $\mu$, which can be related to the lattice and to the mass of the boson field, respectively. In the realm of ultracold atoms, the strength of interactions (in our case the fine-structure constant $\alpha$) is expected to be tunable, which opens the possibility of observing nonperturbative effects. To the best of our knowledge, such models are currently unknown, but we believe that PQED-like gauge theories could be important candidates.

One relevant application of mass generation in PGQED would be the opening of an electronic band gap in graphene in the ultrarelativistic limit, where electrons move at the speed of light and Lorentz symmetry is respected. This limit is more realistic for samples with a very small density of charge carriers, as has been experimentally shown in Ref.~\cite{Geim2011}.  For graphene, the massless Dirac action is naturally obtained as a low-energy approximation of charge carriers. This is in agreement with our model for the matter field, which interacts with the static potential in Eq.~\eqref{Staticpot}. Next, let us discuss how to estimate the main system-dependent constants ($\Lambda,\mu,\alpha$) for this material. For PGQED, the Podolsky parameter $\mu$ can be understood as a phenomenological parameter that decreases the strength of the Coulomb potential, which is a reasonable effect due to the many-body physics of the problem. Indeed, our parameter $\mu$ may provide a phenomenological explanation for the upper limit of the observed gap in graphene $2|M|\leq 0.1$~meV, reported in the experimental work in Ref.~\cite{Geim2011}. In this case, $\Lambda\approx 1$~eV is obtained from the inverse of the lattice parameter $a_L\approx 10^{-10}$m. The fine-structure constant for suspended monolayer graphene is known to be close to $\alpha\approx 2.2$ \cite{CastroNeto2009}. Using these values in Eq.~(\ref{spmcp}) with $2|M|= 0.1$~meV we obtain the value of the Podolsky parameter, namely, $\mu=0.2$~meV. This estimate should be interpreted as an effective low-energy parameter rather than as a fundamental microscopic mass. We may conclude that the parameter $\mu$ in our model is a useful tool because the true many-body interaction among electrons in graphene (and other two-dimensional materials) has not yet been established, but it is expected to have a screened Coulomb potential due to the electromagnetic interaction. For example, large values of the dielectric constant have been reported in Ref.~\cite{Geim2011} for graphene placed above another material. 


\begin{acknowledgements}
The authors thank D. T. Alves, J. A. Helayel, J. D. L. Silva, R. Thibes, and M. Neves for interesting discussions about our work.  C. A. P. C. J. is partially supported by Coordenação de
Aperfeiçoamento de Pessoal de Nível Superior Brasil
(CAPES), finance code 001. L. O. N. and  V. S. A. are partially supported by Conselho Nacional de Desenvolvimento Científico e Tecnológico-Brazil (CNPq-Brazil), process: 408735/2023-6 CNPq/MCTI. V. S. A was also supported by CNPq through a research project Grant process 303779/2026-8.
\end{acknowledgements}

\appendix

\textbf{\section{The dimensional reduction of GQED}
\label{A}}

In this appendix, we derive the dimensional reduction of GQED from (3+1)D to (2+1)D. The Euclidean GQED Lagrangian is given by
\begin{equation}
    \mathcal{L}_P=\frac{1}{4}F_{\mu\nu}F^{\mu\nu}-\frac{a^2}{2}\partial_\nu F^{\mu\nu}\partial^\sigma F_{\mu\sigma}+\frac{\lambda}{2}(\partial_\mu A^\mu)^2,
\end{equation}
where $a$ is the Podolsky parameter, $A_\mu$ is the gauge-field, and $F_{\mu\nu}=\partial_\mu A_\nu-\partial_\nu A_\mu$ is the field-strength tensor. Furthermore, $\lambda$ is the gauge-fixing parameter. After neglecting total derivative terms, we find
\begin{align}\label{lmg}
    \frac{1}{4}F_{\mu\nu}F^{\mu\nu}+\frac{\lambda}{2}(\partial_\mu A^\mu)^2&=-\frac{1}{2}\delta_{\mu\nu}A^\mu\square A^\nu+\nonumber\\
    &+\frac{1}{2}\left(1-\lambda\right)A^\mu\partial_\mu\partial_\nu A^\nu.
\end{align}
For the Podolsky term, we obtain
\begin{align}
    \partial_\nu F^{\mu\nu}\partial^\sigma F_{\mu\sigma}=A^\mu(\partial_\mu\partial_\nu\square)A^\nu-\delta_{\mu\nu}A^\mu(\square\square )A^\nu.
\end{align}
Consequently,
\begin{align}
    \mathcal{L}_P=\frac{1}{2}A^\mu\left[(-\square)(1-a^2\square)\delta_{\mu\nu}+(1-\lambda-a^2\square)\partial_\mu\partial_\nu)\right]A^\nu.
\end{align}

The GQED action can then be written as
\begin{equation}
    \mathcal{L}_P=\frac{1}{2}A^\mu \Delta_{0,\mu\nu}^{-1}A^\nu,
\end{equation}
where $\Delta_{0,\mu\nu}^{-1}$ is the inverse of the gauge-field propagator, given by
\begin{equation}
    \Delta_{0,\mu\nu}^{-1}=(-\square)(1-a^2\square)\delta_{\mu\nu}+(1-\lambda-a^2\square)\partial_\mu\partial_\nu.
\end{equation}
After taking the Fourier transform of this propagator, using $\square\to-p^2$ and $\partial_\mu\to-ip_\mu$, we find
\begin{equation}
    \Delta_{0,\mu\nu}^{-1}(p)=p^2(1+a^2p^2)\delta_{\mu\nu}-(1-\lambda+a^2p^2)p_\mu p_\nu.
\end{equation}

Therefore, the bare gauge-field propagator is given by
\begin{align}
    \Delta^0_{\mu\nu}(p)=\frac{1}{p^2(1+a^2p^2)}\left[\delta_{\mu\nu}-\frac{p_\mu p_\nu}{p^2}\right]+\frac{1}{\lambda}\frac{p_\mu p_\nu}{p^4}.
\end{align}

In order to perform the dimensional reduction, we consider the generating functional associated with the fermionic Green functions. Therefore, we define
\begin{equation}
    Z=Z_0\int DA_\mu D\psi D\bar{\psi}e^{-\int d^{4}x\left[\mathcal{L}_P+\mathcal{L}_F-ej_\mu A^\mu-\bar{\psi}\eta-\bar{\eta}\psi\right]},
\end{equation}
where $\mathcal{L}_F$ is the fermionic Lagrangian, $Z_0$ is a normalization constant, $j_\mu=\bar{\psi}\gamma_\mu\psi$ is the matter current, $\eta$ and $\bar{\eta}$ are sources for the fermionic field. Using the previous results, it follows that
\begin{align}\label{gfoftt}
    Z=&Z_0\int D\psi D\bar{\psi}e^{-\int d^{4}x\left[\mathcal{L}_F-\bar{\psi}\eta-\bar{\eta}\psi\right]}\times\nonumber\\
    &\times\int DA_\mu e^{-\int d^{4}x\left[\frac{1}{2}A^\mu \Delta_{0\mu\nu}^{-1}A^\nu-ej_\mu A^\mu\right]}.
\end{align}
The functional integral over $A_\mu$ can be evaluated using the identity
\begin{equation}
    \int D \varphi \ e^{-\frac{1}{2}\int dx[\varphi \Delta \varphi]+\int dx\varphi J}=\det{\Delta}^{-1/2}e^{\frac{1}{2}\int dx[J \Delta^{-1}J]},
\end{equation}
where $\varphi$ is a real bosonic field, $\Delta$ is an arbitrary matrix, and $J$ is a source term for the field $\varphi$. We denote by $I$ the gauge-field functional integral appearing in Eq.~\eqref{gfoftt}. Completing the square and applying the standard Gaussian functional integration formula, we obtain
\begin{align}
    I&=\int DA_\mu \ e^{-\int d^{4}x\left[\frac{1}{2}A^\mu \Delta_{0\mu\nu}^{-1}A^\nu-ej_\mu A^\mu\right]}\nonumber\\
    &=\int DA_\mu \ e^{-\frac{1}{2}\int d^{4}x\left[A^\mu \Delta_{0\mu\nu}^{-1}A^\nu\right]+\int d^{4}xj_\mu (eA^\mu)}\nonumber\\
    &=\frac{1}{e}\int D(eA_\mu) \ e^{-\frac{1}{2}\int d^{4}x\left[(eA^\mu)\frac{\Delta_{0\mu\nu}^{-1}}{e^2}(eA^\nu)\right]+\int d^{4}xj_\mu (eA^\mu)}\nonumber\\
    &=\frac{1}{e}\det\{\Delta_{0\mu\nu}^{-1}/e^2\}^{-4/2} \ e^{\int d^{4}x d^4y\left[\frac{e^2}{2}j^\mu \Delta_{0\mu\nu}j^\nu\right]},
\end{align}
 where we have performed the simple change of variables $A_{\mu}\rightarrow eA_{\mu}$. Absorbing the resulting constant factor into the normalization constant, $Z_0\rightarrow Z_0'$, the generating functional becomes 
\begin{equation} \label{Zfim}
    Z=Z_0'\int D\bar{\psi} D\psi \ e^{-\int d^{4}x d^4y\left[-\frac{e^2}{2}j^\mu \Delta_{0,\mu\nu}j^\nu+...\right]},
\end{equation}
where the dots in the rhs of Eq.~(\ref{Zfim}) mean terms that depend only on the matter field and, therefore, do not affect the effective gauge interaction derived below.

We now confine the matter current to a plane $x_1-x_2$, which is achieved by taking
\begin{equation} \label{jconf}
    j_\mu=\begin{cases}
    j_\mu^{3D}(x_0,x_1,x_2)\delta(x_3), \ \mu=0,1,2 
    \\
    0, \ \mu=3.
    \end{cases}
\end{equation}
Using Eq.~(\ref{jconf}) in Eq.~(\ref{Zfim}), we find that the effective gauge-field propagator is given by
\begin{align}
    \Delta^0_{\mu\nu}(x,y)\Bigg|_{x_3=y_3=0}&=\int\frac{d^{4}k}{(2\pi)^4}\frac{\delta_{\mu\nu}e^{ik(x-y)}}{k^2(1+a^2k^2)}\Bigg|_{x_3=y_3=0}\nonumber\\
    &=\frac{1}{2}\frac{\delta_{\mu\nu}}{(-\square)^{1/2}}-\frac{1}{2}\frac{\delta_{\mu\nu}}{(-\square+\mu^2)^{1/2}},
\end{align}
where we have used the identity
\begin{equation}
    \frac{1}{k^2(1+a^2k^2)}=\frac{1}{k^2}-\frac{a^2}{1+a^2k^2}=\frac{1}{k^2}-\frac{1}{k^2+\mu^2},
\end{equation}
and $\mu=1/a$, where $\square$ now denotes the d'Alembertian operator in (2+1)D. The same result in Eq.~(\ref{Zfim}) can also be obtained by the following action in (2+1)D, namely,
\begin{equation} \label{PGQEDappa}
    \mathcal{L}_{{\rm{PGQED}}}=\frac{1}{4}\Tilde{F}_{\mu\nu}K_E[\square]\Tilde{F}^{\mu\nu}+\frac{\xi}{2}\partial_\mu\Tilde{A}^{\mu}K[\square]\partial_\nu \Tilde{A}^\nu+\mathcal{L}_F+\mathcal{L}_I,
\end{equation}
where $\xi$ is a gauge-fixing parameter,
$\mathcal{L}_I$ is the standard interaction of QED, and
\begin{equation}\label{opdppqed}
    K_E[\square]=\frac{2(-\square+\mu^2)}{(-\square+\mu^2)(-\square)^{1/2}-(-\square)(-\square+\mu^2)^{1/2}}
\end{equation}
is a pseudo-differential operator. This completes the derivation of PGQED, given by Eq.~(\ref{PGQEDappa}).

\textbf{\section{Numerical solution of nonlinear integral equation}
\label{B}}

In this appendix, we discuss the numerical solutions to the integral equation for $\Sigma(p)$ in both Eq.~(\ref{fgdmmm}) and Eq.~(\ref{SigIntN}), obtained in the rainbow-quenched and rainbow-unquenched approximations, respectively. These solutions are obtained after converting the integral over $k$ into a sum over a discretized variable $k_n\in [p_{{\rm{min}}},\Lambda]$, where $p_{{\rm{min}}}= 10^{-3}$ (in units of $\Lambda$) and $\Lambda=10$. The resulting discretized equations are solved using the repeated trapezoidal quadrature rule and solve a set of algebraic equations for $\Sigma(p)$ for each value of $p$, which is discretized in the same way as $k_n$. In all numerical calculations we use$M=300$ as the number of intervals for $k_n$, hence, $n=1,...,M$ with a separation of approximately $h=(10-10^{-3})/(300-1)\approx0.033$ between $k_n$ and $k_{n+1}$. The same method has been applied in several previous studies of PQED \cite{Alves2013,Nascimento2015}.

\begin{figure}
    \centering
    \includegraphics[width=\linewidth]{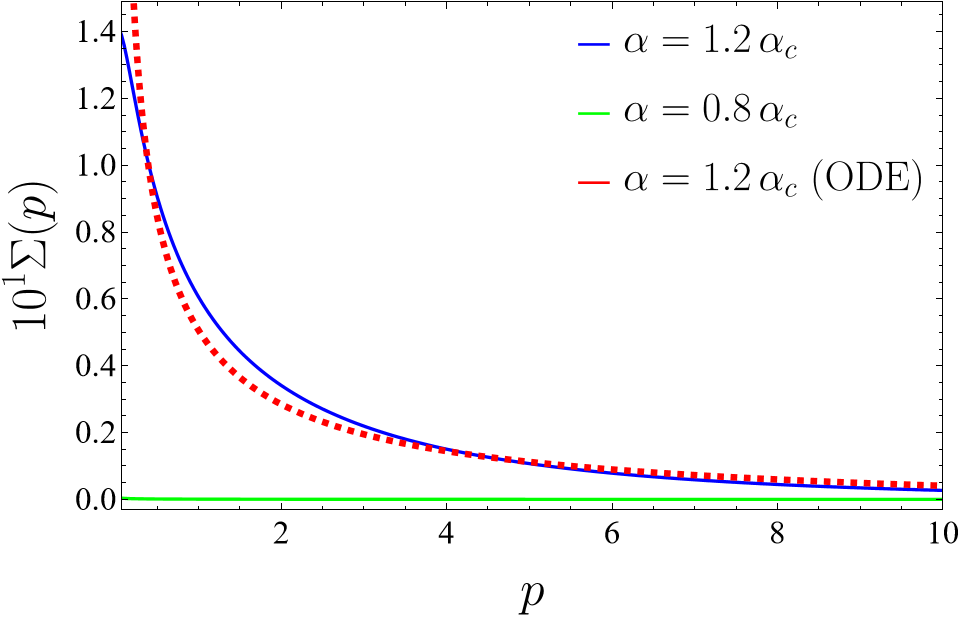}
    \caption{\textit{Numerical and analytical results for PGQED in the rainbow-quenched approximation}. The blue and green lines are the numerical solutions of Eq. \eqref{fgdmmm} as functions of the external momentum $p$ with $\alpha=1.2\alpha_c$ and $\alpha=0.8\alpha_c$, respectively. We consider $\mu=8$ and $\Lambda=10$, which gives $\alpha_c=0.75$ from Eq. \eqref{alcqr}. 
    The red dashed line shows the analytical solution from Eq. \eqref{fgdmee}, with $A_1=0.25-0.32i$ and $A_2=0.25+0.32i$, which is in excellent agreement with the numerical solution for $\alpha=1.2\alpha_c$ in the ultraviolet regime.}
    \label{ppqedrainbowquenched}
\end{figure}

\begin{figure}
    \centering
    \includegraphics[width=\linewidth]{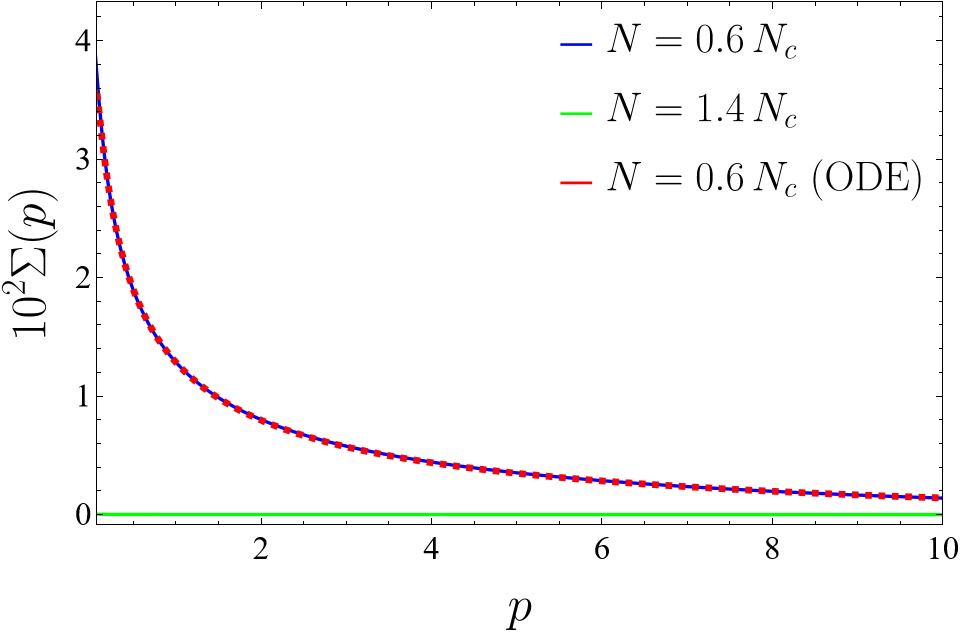}
    \caption{\textit{Numerical and analytical results for PGQED in the rainbow-unquenched approximation}. The blue and green lines are the numerical solutions of Eq. \eqref{aux1} for $N=0.6N_c$ and $N=1.4N_c$, respectively. We consider $\mu=8$, $g=70$ and $\Lambda=10$, which implies from Eq. \eqref{NcN} that $N_c=2.71$. 
    The red dashed line corresponds to the analytical solution from Eq. \eqref{fgdmeee}, with $E_1=0.64-0.19i$ and $E_2=0.64+0.19i$, which is consistent with the numerical solutions for $N=0.6N_c$.}
    \label{ppqedrainbowunquenched}
\end{figure}

In Fig.~\ref{ppqedrainbowquenched}, we show the numerical results for PGQED within the rainbow-quenched approximation. Furthermore, we also plot the analytical solution in Eq. \eqref{fgdmee}, for $\alpha>\alpha_c$, in order to compare the analytical and numerical results. Our numerical results indicate that the mass function is much less than $\Lambda$ for $\alpha<\alpha_c$ and it increases for $\alpha>\alpha_c$, eventually reaching a small fraction of the UV cutoff. The calculated $\alpha_c$ in Eq.~(\ref{alcqr}) yields a reasonable estimate for the critical coupling constant, in particular, it describes the role of the Podolsky parameter $a=1/\mu$.  However, as discussed in Refs.~\cite{Maris1996,Roberts1994}, other approximations may yield different critical coupling constants, for example, by modifying the approximation for the vertex interaction and for $A(p)$. The physical interpretation of $\alpha_c$, however, is the same as the one discussed here. The analytical mass function $\Sigma(p)$ is in excellent agreement to that of the numerical solutions, which is in agreement with previous results for QED3 in Ref.~\cite{Maris1996-2}. Fig.~\ref{ppqedrainbowunquenched} confirms the same conclusions for the rainbow-unquenched approximation.

We now discuss the numerical solution for the wave-function renormalization $A(p)$ within the rainbow-quenched approximation in the symmetric phase $\Sigma(p)=0$. From Eq. (\ref{wfrf}) and Eq. (\ref{sdeff}), after taking the trace over the Dirac matrices, we find
\begin{align}\label{wavefuncA}
    &A(p)=1+\frac{\alpha}{\pi p^2}\int_{0}^{+\infty}\frac{d k}{A(k)}\int_{0}^{\pi} d{\theta}\sin{\theta}
    \times\nonumber\\
    &\times\left[\frac{1}{\sqrt{p^2+k^2-2pk\cos{\theta}}}-\frac{1}{\sqrt{p^2+k^2-2pk\cos{\theta}+\mu^2}}\right]\times\nonumber\\
    &\times\frac{
    [p^3 k \cos{\theta}-k^2p^2\cos^2{\theta}-p^2 k^2 +k^3p \cos{\theta}]}{p^2+k^2-2pk\cos{\theta}}.
\end{align}

\begin{figure}
    \centering
    \includegraphics[width=\linewidth]{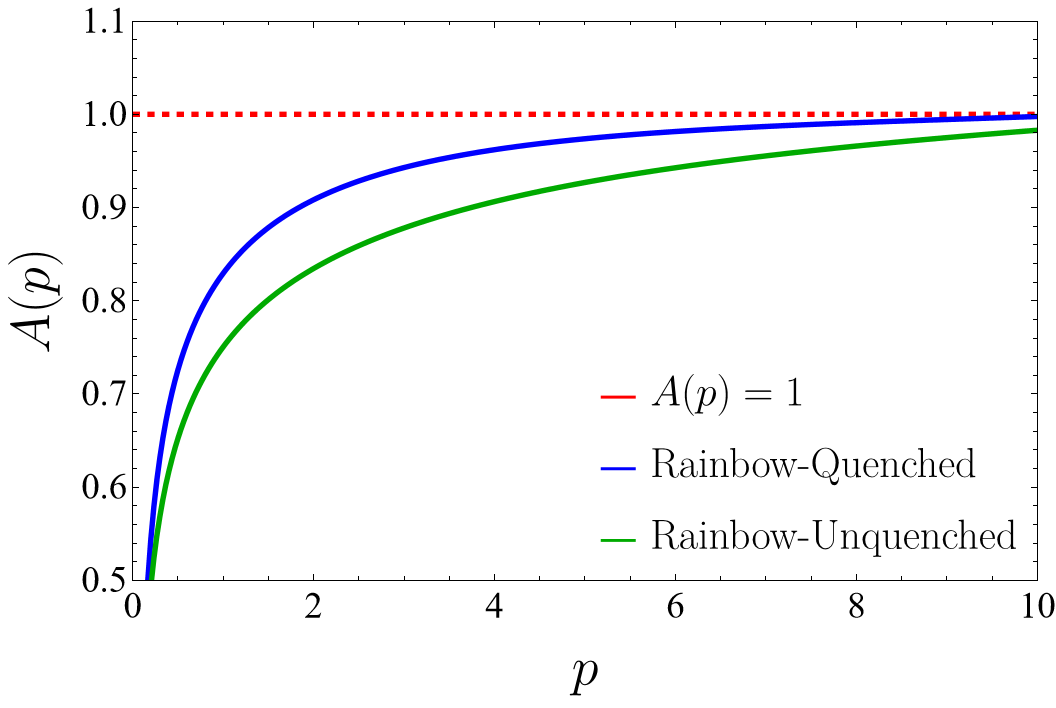}
    \caption{\textit{Numerical results for the wave-function renormalization $A(p)$}. The blue line corresponds to the rainbow-quenched approximation, obtained as the numerical solution of Eq.~\eqref{wavefuncA} in the symmetric phase $\Sigma(p)=0$, for $\mu=10$, $\alpha_c=0.61$, and $\alpha=1.2\alpha_c$. The dark green line represents the rainbow-unquenched case, computed for $g=100$, $\mu=10$, $N_c=3$, and $N_f=0.7N_c$. In both cases, $A(p)\rightarrow 1$ as $p\rightarrow \Lambda$.}
    \label{wavefuncrenormfunc}
\end{figure}

From Fig.~\ref{wavefuncrenormfunc}, we conclude that $A(p)\approx 1$ provides a reasonable leading-order approximation in the large-external-momentum regime within the present truncation. In this context, we note that the rainbow-unquenched equation for $A(p)$ is structurally analogous to Eq.~\eqref{wavefuncA}, differing only through the replacement of the interaction kernel and the substitution $e^2 \rightarrow g/N$. We emphasize, however, that sizable deviations are expected in the deep infrared and in treatments including improved vertex constructions.

\textbf{\section{The anisotropic case}
\label{C}}

In this appendix, we present the derivation of the anisotropic gap equation and its critical coupling constant, $\alpha_c^*$, within the rainbow–quenched approximation. Anisotropy arises when considering systems in which the quasiparticles propagate with a Fermi velocity $v_F$ that is different from the speed of light $c$. To incorporate the explicit Lorentz-symmetry breaking characteristic of Dirac materials, we replace the spatial derivatives by  $v_F\partial_i$, leading to the anisotropic Dirac operator $i\gamma_\mu\partial^\mu \to i\gamma_0\partial^0 + i v_F \gamma_i \partial^i$ , as commonly adopted in effective descriptions of graphene \cite{Marino2017}. We emphasize that the anisotropic analysis presented here is intended to explore qualitative trends associated with Lorentz-symmetry breaking, rather than to provide a quantitative description of specific materials. We then minimally couple the fermion field to the PGQED gauge-field and obtain
\begin{align}
    \mathcal{L}_{APGQED} =& \frac{1}{4} \left(\mathbf{E} K_E[\square] \mathbf{E} 
    - \mathbf{B} K_E[\square] \mathbf{B} \right)\nonumber\\
    &+ \bar{\psi} \left(i\gamma^{0}\partial_0 + i v_F \gamma^{i} \partial_i - mc^2 \right) \psi \nonumber \\
    & -\, e\, \bar{\psi} \gamma^0 \psi A_0 
    - \frac{e v_F}{c} \bar{\psi} \gamma^i \psi A_i \nonumber \\
    &+ \frac{\xi}{2} A_\mu \partial^\mu K_E[\square] \partial^\nu A_\nu,
\end{align}
where $K_E[\square]$ is given by Eq.~\eqref{KernelRGQED}. From this, we obtain the Feynman rule for the gauge-field propagator, namely,
\begin{equation}
    \Delta_{\mu\nu}^{0}(p)=\left[\frac{c}{2\sqrt{p^2}}-\frac{c}{2\sqrt{p^2+\mu^2c^4}}\right]P_{\mu\nu},
\end{equation}
in Euclidean space. The bare fermion propagator is 
\begin{equation}
    S_F^0=\frac{1}{-\gamma^0p_0-v_F\gamma^ip_i+mc^2}=\frac{\gamma^0p_0+v_F\gamma^ip_i+mc^2}{p_0^2+v_F^2\mathbf{p}^2+m^2c^4},
\end{equation}
where $m$ is the bare fermion mass and the interaction vertex is
\begin{equation}
    \Gamma^\mu_0=e\left(\gamma^0,\frac{v_F}{c}\gamma^i\right).
\end{equation}
The full electron propagator can be decomposed in terms of two scalar functions, i.e.,
\begin{equation}
    S^{-1}_{F}(p)=-\gamma^0p_0-B(p)v_F\gamma^ip_i+\Sigma(p)c^2,
\end{equation}
where $B(p)$ is the fermion wave-function renormalization and $\Sigma(p)$ is the mass function.

As discussed previously, the wave-function renormalization satisfies $B(p)=1+\mathcal{O}(e^2)$. Consequently, the approximation $B(p)\approx1$ is adopted in leading order within the present truncation. We expect significant deviations from this behavior deep in the infrared regime, a detailed investigation of which is left for future work. Since $v_F/c\ll1$, the spatial part of the interaction vertex is suppressed, thus we consider $\Gamma^{\mu}\rightarrow \gamma^0$, so that the temporal component dominates. Furthermore, dynamical mass generation is governed by the static interaction kernel, allowing us to set $p_0=0$ in the gauge propagator, as is standard in PQED and QED$_3$ gap-equation analyzes. We also consider massless fermions, $m=0$, which is appropriate for Dirac materials such as graphene, where the bare fermion mass vanishes. Using the two-dimensional integration measure $d^2k=kdkd\varphi$ together with $|\mathbf{p}-\mathbf{k}|^2=p^2+k^2-2pk\cos{\varphi}$, one obtains the following 
\begin{align}
    \Sigma(p)&=\pi v_F\alpha\left(1+\frac{v_F^2}{c^2}\right)\int_{0}^{\Lambda}\frac{d{k}}{2\pi}\frac{k\Sigma(k)}{\sqrt{v_F^2\mathbf{k}^2+\Sigma(k)^2c^4}}\nonumber\times\\
    &\times[Q(k,p)-R(k,p,\mu)],
\end{align}
where
\begin{equation}
    Q(k,p)=\frac{K\left(-\frac{4 k p}{(k-p)^2}\right)}{\pi\sqrt{(k-p)^2}}+\frac{K\left(\frac{4 k p}{(k+p)^2}\right)}{\pi\sqrt{(k+p)^2}}
\end{equation}
and
\begin{equation}
    R(k,p,\mu)=\frac{K\left(-\frac{4 k p}{(k-p)^2+\mu^2c^4}\right)}{\pi\sqrt{(k-p)^2+\mu^2c^4}}+\frac{K\left(\frac{4 k p}{(k+p)^2+\mu ^2c^4}\right)}{\pi\sqrt{(k+p)^2+\mu^2c^4}},
\end{equation}
where $K(x)$ is the complete elliptic integral of the first kind. As in isotropic theory ($v_F=c$), we introduce the ultraviolet cutoff $\Lambda$. Retaining the leading-order terms in the limits $p\gg k$ and $k\gg p$, using the same steps as in the isotropic case, we find an integral equation for the mass function, given by
\begin{align}
    \frac{d}{dp}&\left\{H(p,\mu)p^2\Sigma'(p)\right\}+\nonumber\\
    &+\frac{v_F\alpha}{2}\left(1+\frac{v_F^2}{c^2}\right)\frac{p\Sigma(p)}{\sqrt{v_F^2\mathbf{p}^2+\Sigma(p)^2c^4}}=0.
\end{align}
where $H(p,\mu)$ is given by Eq.~\eqref{funcH}. Here, we consider only the simplest case where $c^2\Sigma^2(p)\ll v_F^2p^2$ and the same approximation used previously $H(p,\mu)p^2\to H(\Lambda,\mu)p^2$, which is enough to describe the mass function as long as $\Lambda>p$. This assumption restricts the present analysis to the weak-gap regime and is not expected to hold deep in the infrared, where nonperturbative effects become dominant. Furthermore, higher-order terms in $v_F/c$ are retained only for qualitative comparison and should not be interpreted as a controlled expansion. With these approximations in mind, we obtain a Cauchy-Euler differential equation, namely,
\begin{equation}\label{fgdmaniso}
    p^2\Sigma''(p)+2p\Sigma'(p)+\frac{\alpha}{2 H(\Lambda,\mu)}\left(1+\frac{v_F^2}{c^2}\right)\Sigma(p)=0.
\end{equation}

The solution of Eq. \eqref{fgdmaniso} allows us to find the critical coupling constant,
\begin{equation}\label{alcqraniso}
    \alpha_c^*=\frac{H(\Lambda,\mu)}{2\left(1+\frac{v_F^2}{c^2}\right)}=\frac{1}{2\left(1+\frac{v_F^2}{c^2}\right)}\left[\frac{1}{1-\frac{\Lambda^3}{( \Lambda^2+\mu^2c^4)^{3/2}}}\right].
\end{equation}
For $\mu \gg \Lambda$ ($a\Lambda \ll 1$), we obtain $\alpha_c^* \to \frac{1}{2(1+v_F^2/c^2)} \approx 0.49$ for $v_F \approx c/300$, a result that was originally obtained in Ref.~\cite{Reginaldo2015} for PQED. As expected, increasing the Fermi velocity decreases the critical value $\alpha_c^*$ predicted by the present truncation, thereby favoring chiral symmetry breaking. Moreover, as in the isotropic case, the Podolsky parameter tends to move the theory to the symmetric phase. These results modify the critical coupling constant discussed in Sec. VI, now depending on $v_F$ as shown in Eq.~\eqref{alcqraniso}. We stress that the critical coupling constant obtained from Eq.~\eqref{fgdmaniso} should be interpreted as a qualitative indicator of the onset of dynamical mass generation within the present truncation. More accurate quantitative predictions are expected to require improved vertex constructions together with a fully self-consistent treatment of the wave-function renormalization.

Finally, we briefly comment on the renormalization of the Fermi velocity in graphene. In Ref.~\cite{Geim2011}, the authors showed that it increases as the electron density decreases in graphene. To account for this behavior, they used a logarithmic expression for the renormalized Fermi velocity, obtained in Ref.~\cite{Gonzlez1994}. This expression diverges at zero density and is derived assuming a static Coulomb interaction. When a full electromagnetic interaction is considered within the PQED framework, one finds that the Fermi velocity tends toward the speed of light in this limit \cite{Marino2017}, which is an important test of consistency of the theory. In the present analysis, however, we have shown that the dominant effects of anisotropy on dynamical mass generation are primarily encoded in the modified critical coupling constant, whereas the underlying mechanism of chiral symmetry breaking remains unchanged.

\vspace{1 cm}


\printbibliography

\end{document}